\documentclass[pagebackref=true]{raa}   
\usepackage[T1]{fontenc}

\usepackage{graphicx,times} 
\usepackage{float}
\usepackage{natbib}
\usepackage{amssymb,amsmath}
\bibpunct{(}{)}{;}{a}{}{,}
\usepackage{hyperref}
\usepackage{siunitx}
\usepackage{soul,xcolor}
\setstcolor{red}

\usepackage{tikz}
\usetikzlibrary{shapes, arrows}
\usepackage{caption}

\begin{document}

  \title{Spectroscopic study of late-type emission-line stars using the data from LAMOST DR6}

   \volnopage{Vol.0 (20xx) No.0, 000--000}      
   \setcounter{page}{1}          

   \author{D. Edwin \inst{1} \and 
        Blesson Mathew \inst{1} \and 
        B. Shridharan \inst{1,2} \and 
        Vineeth Valsan \inst{1} \and 
        S. Nidhi \inst{1,3} \and 
        Suman Bhattacharyya \inst{1} \and 
        Sreeja S. Kartha \inst{1} \and 
        T. Robin \inst{1} 
        }

   \institute{Department of Physics and Electronics, CHRIST (Deemed to be University), Hosur Main Road, Bangalore, India \\
    {\it blesson.mathew@christuniversity.in, edwin.jv@res.christuniversity.in}\\
   \and Department of Astronomy and Astrophysics, Tata Institute of Fundamental Research, Mumbai, 400005, India\\
   \and The Oxford College of Science, 17th, 32, 19th Main road, Sector 4, HSR Layout, Bengaluru, Karnataka, 560102\\
\vs\no
   {\small Received 20xx month day; accepted 20xx month day}}

\abstract{ Low-mass emission-line stars belong to various evolutionary stages, from pre-main-sequence young stars to evolved stars. In this work, we present a catalog of late-type (F0 to M9) emission-line stars from the LAMOST Data Release 6. Using the \texttt{scipy} package, we created a Python code that finds the emission peak at $H\alpha$ in all late-type stellar spectra. A dataset of 38,152 late-type emission-line stars was obtained after a rigorous examination of the photometric quality flags and the signal-to-noise ratio of the spectra. Adopting well-known photometric and spectroscopic methods, we classified our sample into 438 infrared excess sources, 4,669 post-main-sequence candidates, 9,718 Fe/Ge/Ke sources, and 23,264 dMe sources. From a cross-match with known databases, we found that 29,222 sources, comprising 65 IR excess sources, 7,899 Fe/Ge/Ke stars, 17,533 dMe stars, and 3,725 PtMS candidates, are new detections. We measured the equivalent width of the major emission lines observed in the spectra of our sample of emission-line stars. Furthermore, the trend observed in the line strengths of major emission lines over the entire late-type spectral range is analyzed. We further classified the sample into 4 groups based on the presence of Hydrogen and Calcium emission lines. This work presents a large dataset of late-type emission-line stars, which can be used to study active phenomena in late-type stars.
}

   \authorrunning{Edwin et al. }            
   \titlerunning{Spectroscopic study of LDR6 LELS}  

   \maketitle

%
%
\section{Introduction}           
\label{sect:intro}

The emission of $H\alpha$ is frequently observed in active pre- and post-main-sequence stars, as well as in binary stars. The investigation of stars exhibiting $H\alpha$ emission holds considerable physical significance. Historically, the study of these stars commenced in the early 1940s where \cite{merrill1933catalogue} published their work on spectroscopic observations of $H\alpha$ emission stars in the northern Milky Way, which sparked interest in emission-line stars (ELS). Following this, \cite{bidelman1954catalogue} made a catalog of stars later than spectral type B, which showed emission in $H\alpha$ and $Ca~{\textsc{ii}}$ H $\&$ K. \cite{wray1966new} and \cite{henize1976observations} conducted research on $H\alpha$ emission stars in the southern Milky Way, which further helped in expanding our understanding of ELS. In the 1980s, multiple surveys such as \cite{macconnell1981discoveries}, \cite{stephenson1986new}, and \cite{wiramihardja1989survey} extensively studied ELS in different parts of our galaxy. In subsequent years, ELS were also surveyed and identified beyond our galaxy, notably from the Large Magellanic Cloud (LMC; \citealt{bohannan1987spectrophotometry}) and Small Magellanic Clouds (SMC; \citealt{meyssonnier1993new,le2005h,martayan2010slitless}). 

The advent of the Sloan Digital Sky Survey (SDSS;\cite{york2000sloan}) has contributed to the field of stellar spectroscopy and improved the understanding of various active phenomena in stars. Also, Isaac Newton Telescope (INT) Photometric H-$\alpha$ Survey (IPHAS; \cite{drew2005int}), a program that used broad-bands \emph{r} and \emph{i}, and narrow-band $H\alpha$ filters, studied the northern plane of our galaxy. One of the specific goals of this survey program was to study the YSOs in our galaxy \citep{barentsen2014second}. The Gaia-ESO survey \citep{traven2015gaia}, which studied the spectra of a number of active ELS also improved the list of ELS. These build a great foundation for the upcoming studies and surveys on ELS in our galaxy. Furthermore, RAdial Velocity Experiment (RAVE;\citealt{steinmetz2002rave}), an all-sky spectroscopic survey to measure stellar parameters like radial velocity, metallicity, and abundance ratios of 50 million stars in our galaxy contributed much to the understanding of ELS research.

The study of ELS in the low-mass regime is critical due to the occurrence of a diverse range of active phenomena in these sources. Late-type emission-line stars (LELS) are low-mass ($<$ 3 $M_{\odot}$) emission-line stars (ELS) belonging to spectral types F0–M9. LELS can be broadly classified into infrared excess sources, Fe/Ge/Ke/dMe (main-sequence) stars, and post-main-sequence (PtMS) candidates. Classical T Tauri stars (CTTS) are young stellar objects (YSO) that undergo accretion, with $H\alpha$ emission originating primarily from the accretion columns and the circumstellar disk \citep{kurosawa2005formation}. These stars exhibit significant infrared excess. In main-sequence Ge/Ke/dMe stars, the $H\alpha$ emission occurs from the chromosphere of the star, which is usually attributed to the large-scale magnetic fields observed in low-mass stars \citep{hall2008stellar}. In PtMS F- and G-type stars, $H\alpha$ originates from the expanding outer shell and ejected material \citep{kogure2010astrophysics}. In stars like Mira variables, the origin of the emission lines is due to the pulsations leading to the generation of shock waves, which results in intense emission lines in their spectra \citep{wood1979pulsation,gillet1983shock,kogure2010astrophysics}. Hence, the primary objective of this work is to classify LELS into distinct evolutionary stages. 


The major emission lines observed in the optical spectra of late-type stars include $Ca~{\textsc{ii}}$ H $\&$ K, $H\alpha$, and the $Ca~{\textsc{ii}}$ triplet (CaIRT). Most of these lines are often linked with chromospheric activity in low mass stars \citep{cincunegui2007halpha,smith2011some}. Other factors that influence the strength and profile shape of the emission lines are due to stellar rotation, age of the star, wind, and magnetic field. Multiple photometric \citep{radick1983photometric, herbst1989h, lockwood2007patterns} and spectroscopic \citep{gizis2002palomar, lopez2010high, suarez2015rotation} studies provided us with good insights on different groups of LELS in our galaxy. Notably, utilizing SDSS spectra, \cite{west2004spectroscopic} studied the magnetic properties of cool stars, where they used emission in $H\alpha$ as the primary indicator of activity. Also, RAVE studies on 38,000 chromospherically active solar-type stars in the solar neighborhood was carried out by \cite{vzerjal2017chromospherically}.

Large Sky Area Multi-Object Fibre Spectroscopic Telescope (LAMOST) has provided a huge spectral database for the scientific community, containing more than 10 million low-resolution spectra of stars, galaxies, and QSOs. There are several works that have used the LAMOST database to study different types of LELS, such as Mira variables \citep{yao2017mira}, M dwarfs \citep{yi2014m,guo2015m,chang2017lamost} and M-giants \citep{liu2014k,zhong2015m,li2019detecting}. The present study complements the work of \cite{shridharan2021discovery}, where they created a catalog of early-type ELS using the spectra available in LAMOST data archive. We queried the entire LAMOST spectral database for sources classified as F, G, K, and M spectral types and identified the spectra showing the $H\alpha$ emission line. We then classify our sample into Fe/Ge/Ke/dMe stars, PtMS candidates, and YSOs on the basis of their photometric and spectroscopic features. It improves the existing sample of LELS, in turn opening the gateway for studying interesting low-mass sources of our galaxy and improving our understanding on the chromospheric, rotational, magnetic, and all other kinds of activity in these sources. 

In this work, we catalog 38,152 LELS into their various categories and study its spectral characteristics. In Section \ref{sec:method}, we describe the data collection method utilized for our work. In Section \ref{sec:classification}, we provide a brief description on different classes of LELS and the classification criteria utilized in classifying our sample. In Section \ref{sec:result}, the results of this work that contain different types of LELS were separated, and their properties and characteristics were described. 


\section{Data collection \& Compilation} 
\label{sec:method}

\subsection{Identification of late-type stars from LAMOST DR6 spectral database}

LAMOST is a reflecting Schmidt telescope \citep{zhao2012lamost}, observing a field of view (FoV) of 20 $deg^{2}$ \citep{sun2021catalog} in the sky. By July 2018, LAMOST had released its sixth data release (DR6), observing 9,911,337 low-resolution spectrum (LRS). They were subdivided into four categories, namely ``STAR", ``GALAXY", ``QUASAR", and ``unknown" spectra. Among the 4 categories, over 90\% of the LRS belong to the ``STAR" category. A typical LRS contains calibrated wavelength on the x-axis and flux on the y-axis, with a wavelength range of 3690 $\mathring{A}$ - 9100 $\mathring{A}$ and a resolution of 1800 at 5500 $\mathring{A}$ \citep{zhao2012lamost,yan2022overview}. The survey spans over a declination range of $=$ -10$^{\circ}$ to +90$^{\circ}$ \citep{deng2012lamost}. 

For the present work, we have restricted the analysis to a sample of F, G, K, and M spectral types. Hence, we queried for these stars in the LAMOST DR6 (LDR6) catalog and obtained more than 1.7 million star spectra belonging to F type, $\sim$ 2.9 million G-type spectra, $\sim$ 1 million K-type spectra, and $\sim$ 0.55 million M-type spectra. The general catalog of LDR6 contains 37 columns worth of data including the stars' positions, error estimates, identifiers, and magnitudes in \emph{u}, \emph{g}, \emph{r}, \emph{i}, and \emph{z} bands, redshift, etc \citep{wu2014automatic}. The LRS A, F, G, K type star catalog has, along with the general catalog data, information on effective temperature ($T_{eff}$), surface gravity (log($g$)), metallicity [Fe/H], heliocentric radial velocity (rv), and its corresponding errors. These parameters are estimated by the LAMOST parameter pipeline (LASP) \citep{wu2014automatic}. The LRS M-type star catalog has estimated the parameters mentioned in the previous catalog using the cross-correlation method. Furthermore, the catalog provides additional information, which includes $H\alpha$ equivalent width (EW) and the line indices (TiO, CaH, CaOH, Na) estimated using the HAMMER software \citep{liu2014k}. 

\subsection{Selection of late-type emission-line stars}

Initially, 8,613,834 F-, G-, K-, and M-type spectra labeled as ”STAR” were queried from the LDR6 $\footnote{\url{www.lamost.org/dr6}}$ and the spectra were continuum normalized using \texttt{laspec} package \footnote{\url{ https://github.com/hypergravity/laspec}} \citep{zhang2020deriving}. Using the $\texttt{scipy.signal.find peaks}\footnote{\url{https://docs.scipy.org/doc/scipy/reference/generated/scipy.signal.find_peaks.html}}$ parameter, a Python routine was developed to inspect for the presence of $H\alpha$ emission in the sample \citep{shridharan2021discovery}. The parameter ``width” was utilized to avoid false emission sources, only the sources with width greater than 3 sampling points were considered as detection. This avoided narrow peaks caused by instrumental noise or defects. If the spectra does not show $H\alpha$ in emission, they are removed from further analysis. As a result, we obtained 80,860 spectra with the $H\alpha$ emission line within 3 $\si{\angstrom}$ of 6563 $\si{\angstrom}$. Only stars having signal-to-noise ratio in $\texttt{r}$ band (SNR$_{r}$) greater than 10 are further considered for the study, and the rest are eliminated, leaving us with a sample of 58,508 LELS spectra. We further separated the sources having LELS spectra with SNR(x)=-9999 (where x represents $\emph{u}$, $\emph{g}$, $\emph{r}$, $\emph{i}$, and $\emph{z}$ bands). Those spectra were considered ``BAD spectra" upon visual check and removed from this study, which left us with a sample of 56,291 spectra belonging to 48,657 unique sources. Furthermore, we utilized \texttt{line\_index} function of \texttt{laspec} package to estimate the equivalent width of major emission lines for all the sources in our study.

\subsubsection{Cross-match with photometric catalogs}

In order to characterise the identified $H\alpha$ source, we obtain photometric values from various optical/IR all-sky surveys such as Two Micron All-Sky Survey (2MASS) \citep{skrutskie2006two}, Gaia \citep{prusti2016gaia, vallenari2023gaia}, and Wide-field Infrared Survey Explorer (WISE) \citep{cutri2012wise}. To avoid contamination from multiple sources, we used a search radius of 3\arcsec and selected the closest sources to the observed coordinates. Crossmatching with 2MASS and using a quality cut of `AAA' reduced the sample size to 45,376. This sample was then crossmatched with the Gaia DR3 to obtain astrometry and optical photometry, which further reduced the sample to 45,268. The sample further reduced to 44,520 sources on crossmatching with WISE catalog. Of this, we considered only 43,329 LELS with distance value available from \cite{bailer2021estimating}. The line-of-sight extinction values for our sample were obtained using the probabilistic Galactic dustmaps model, Bayestar 2019 \citep{2018JOSS....3..695M}.

\section{Late-type emission-line stars: Classes and classification schemes} \label{sec:classification}

In this section, we provide an overview of LELS along with its literature and the classification criteria employed in segregating them to different classes. 

\subsection{Classes of LELS} \label{subsec:classes}

We categorize LELS into four classes: 

\begin{itemize}
    \item Infrared (IR) excess sources
    \item PtMS sources
    \item Fe/Ge/Ke sources
    \item dMe sources
\end{itemize}

IR excess sources are characterized by the presence of circumstellar disks or envelopes composed of dust and gas. These structures absorb stellar radiation and subsequently emit it at longer wavelengths, resulting in an excess of infrared radiation in their spectral energy distribution (SED). Accretion disks are commonly found around young stellar objects, such as protostars, T Tauri stars, and Herbig Ae/Be stars \citep{groenewegen2012infrared,Hartmann2016ARA&A..54..135H,brittain2023SSRv..219....7B}. Some evolved stars, including those classified as asymptotic giant branch (AGB) stars and red giants, also exhibit IR excess \citep{groenewegen2012infrared}. The accretion disk and mass loss are the major reasons for the formation of emission lines in these objects \citep{muzerolle1998magnetospheric}. 

PtMS stars, ranging in spectral types from F0 to M9, represent stars that have transitioned beyond their main-sequence phase, generally classified into red giants(RGBs), horizontal branch stars (HBs), and asymptotic giant branch stars (AGBs), based on their evolutionary phase. Activity in these sources is due to phenomena such as stellar winds, mass ejections, and pulsations. One of the well-studied groups of PtMS is the Mira variables, where activity happens in the form of long, periodic pulsations \citep{wood1979pulsation,yao2017mira}. In stars such as RGBs and K supergiants, variability in $H\alpha$ has been observed, which has been directly correlated with the mass loss \citep{vasu1982study,cacciari1983survey}.


dMe stars exhibit significant magnetic field due to the presence of convective outer layers and rapid rotation \citep{hall2008stellar}. The magnetic fields are generated by dynamo processes, where the convective material is moved within the star's convective zone, which amplifies the magnetic field. These will lead to star spots, flares, and prominences in low-mass sources. Metallic lines such as Fe, Ca, and Na, which are sensitive to magnetic field, can be utilized to study magnetic field effects such as Zeeman splitting and Stokes broadening \citep{morin2012magnetic}. Higher resolution spectra of these sources will help accurately study the effect of magnetic field in these sources. Red dwarf emission stars (dMe stars) exhibit emission in $H\alpha$ and/or Ca II H $\&$ K, both of which serve as indicators of chromospheric activity. dMe stars were classified into weak dMe stars (0.1 $\si{\angstrom}$ $\leq$ EW($H\alpha$) $\leq$ 0.5 $\si{\angstrom}$) and strong dMe stars (EW($H\alpha$) $<$ 0 $\si{\angstrom}$) \citep{kogure2010astrophysics}. Furthermore, \cite{joy1974spectral} classified dMe stars from dMs photometrically, creating a statistics of the frequency of occurrence of emission in $H\alpha$ across different classes of M stars for the 426 M-type stars they observed. A spectroscopic investigation of the chromospheric activities on dMe stars were conducted \citep{cram1989chromospheres,robinson1990chromospheric}, whereby the intensity and excess emission in a few of the observed emission lines are modeled \citep{young1989study}. Studies on activity indices of dMe stars along with its relationship with age was also carried out \citep{vaughan1980survey,pace2009investigation}. There is a special class of dMe stars known as flare stars or UV Ceti type stars, which show emission in $H\alpha$, Ca II H $\&$ K, and in some cases, He and Na in emission during their quiescent state. Most of these lines get enhanced during flare state \citep{eason1992spectroscopic}. Flares in M dwarfs are explosive events that occur as a result of magnetic reconnection in the atmosphere. A sample of 49 M dwarfs were studied for different spectra observed over a period of time using SDSS \citep{hilton2010m}. Further, 49 more flaring M dwarfs were detected using GALEX data by \cite{welsh2007detection}. The Far-UV activity of the M-dwarfs was studied by \cite{france2018far}. Flares in 480 M dwarfs were identified, and \cite{martinez2020catalog} have created a catalog. The LAMOST survey offers supplementary spectral information for M-type stars. This information comprises the equivalent widths and errors of $H\alpha$ as well as the spectral indices of other molecular bands, including CaH and TiO$_{5}$. The line strengths of TiO versus CaH are used as a popular method in the separation of late-type giants from dwarfs \citep{mould1977band,zhong2015m,yi2019efficient}. The CaH molecular band acts as an indicator of surface gravity, whereas TiO acts as a temperature indicator. 


Fe/Ge/Ke sources are F-, G-, and K-type main-sequence stars or dwarfs showing a considerable amount of chromospheric activity leading to the formation of the emission of lines such as $H\alpha$ and $Ca~{\textsc{ii}}$ H $\&$ K. In the Sun, non-radiative heating is closely linked to the magnetic field, which drives chromospheric activities such as sunspots and flares, ultimately resulting in the emission of $Ca~{\textsc{ii}}$ \citep{robinson1990chromospheric}. Young stars such as weak-lined T Tauri stars will also be included in this group due to the absence of IR excess \citep{gras2005weak,padgett2006spitzer}.

\subsection{Classification Criteria} \label{subsec:classification}

In this section, we have elaborated the classification criteria employed for the separation of LAMOST LELS into different categories.

\subsubsection{Identifying sources with infrared excess} \label{sub_sec:tts_cut}

\begin{figure}[H]
\centering
  \includegraphics[width=1.0\textwidth]{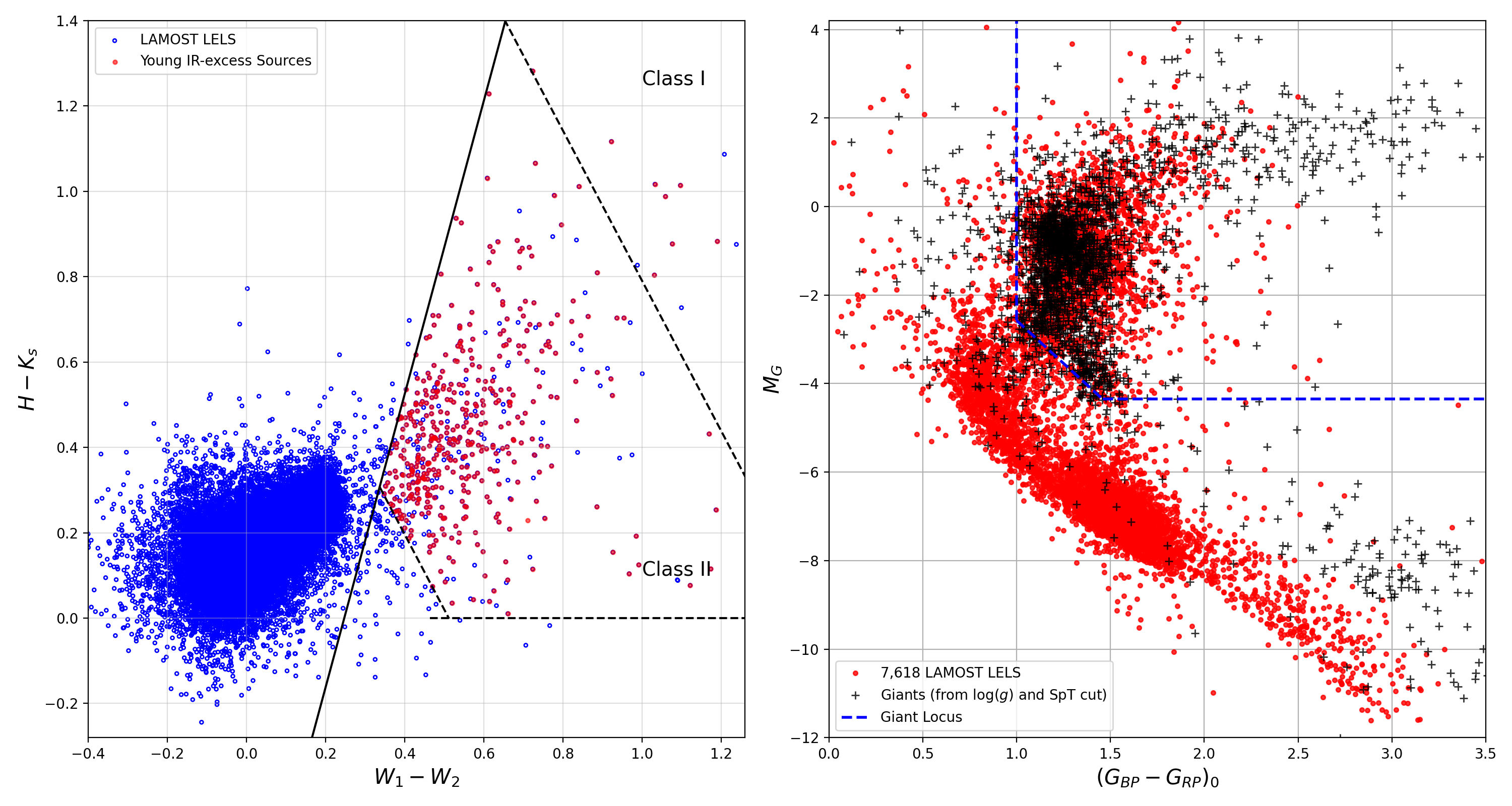}
  {\phantomsubcaption\label{fig:tts_cut}%
     \phantomsubcaption\label{fig:giant_cut}}%
  \caption{(a) The 2MASS-WISE CCDm for the selection of young IR-excess candidates from LAMOST DR6 is displayed in the left panel. The blue dots represent the entire sample of LELS, and the red dots represent IR-excess sources. The dotted and solid lines represent the selection region for Class I and Class II sources defined in \cite{koenig2014classification}. The blue dots in the Class II region represent stars with spectral types `gM'. (b) Gaia CMD for the selection of LAMOST PtMS candidates is displayed in the right panel. The red dots in the plot represent 7,618 LAMOST LELS, whose selection is mentioned in Section \ref{sub_sec:giant_cut}. The black crosses represent PtMS candidates segregated using the log($g$) and the spectral type (SpT) cuts. The blue dotted line represents the region dominated by PtMS candidates.}
\end{figure}

We made use of the 2MASS-WISE color-color diagram (CCDm) for separating young stars showing IR excess. Generally, YSOs are classified into Class {\sc I}, Class {\sc II} and Class {\sc III} sources based on the continuum slope in the IR region of the spectral energy distribution (SED) of the star \citep{1987IAUS..115....1L}. Class {\sc I} sources are protostellar candidates that are deeply embedded in the molecular clouds, and Class {\sc II} sources are classified as T Tauri stars that have almost dispersed its envelope, but still have an accretion disk around it \citep{1993ApJ...406..122A}. Class {\sc III} sources are evolved YSOs where disk dissipation is underway. \cite{koenig2014classification} introduced the criteria to classify the YSOs into various classes using 2MASS-WISE CCD. In Figure \ref{fig:tts_cut}, we have reproduced the YSO classification using the criteria mentioned in \cite{koenig2014classification} and is listed below.

For Class II sources,

\begin{align*}
    H - K_{s} &> 0   \\and \\    
    H - K_{s} &> -1.76*(W_{1} - W_{2}) + 0.9    
    \\and \\
    H - K_{s} &< (0.55/0.16) * (W_{1} - W_{2}) - 0.85    \\and \\
    W_{1} \leq 13 
\end{align*}

and for Class I sources,

\begin{align*}
    H - K_{s} &> -1.76 * (W_{1} - W_{2}) + 2.55   
\end{align*}

Given that the primary focus of this section is to segregate young sources, we utilized the LAMOST spectral type classification. Consequently, stars identified as giant-M-type (``gM") by LAMOST were removed from the sample. This step effectively reduces the contamination of evolved stars in our dataset.

After applying this additional criteria, we segregated a sample of 438 IR excess sources.

\begin{figure}
\centering
  \includegraphics[width=0.7\linewidth]{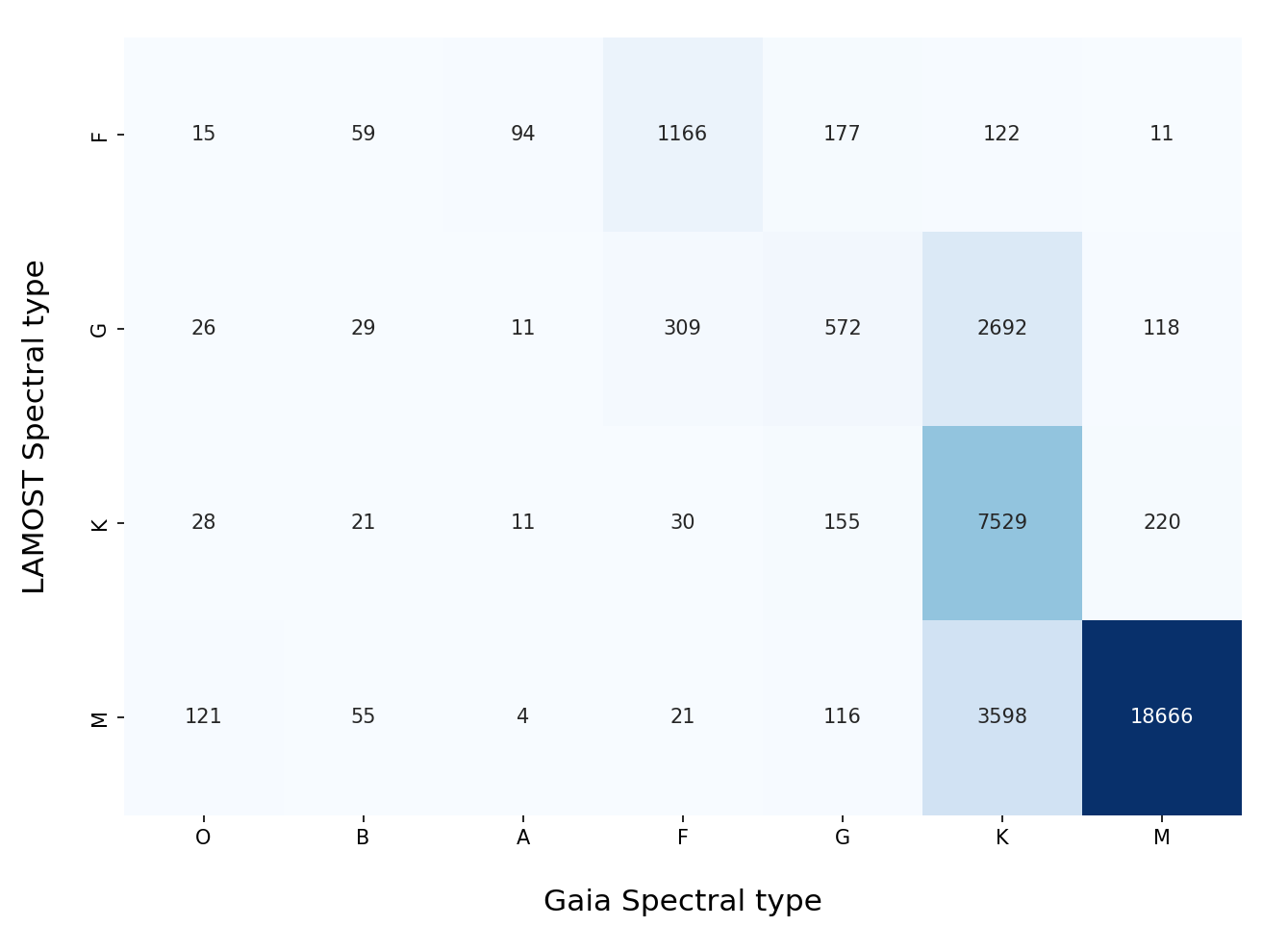}  
  \caption{The number distribution plot of the LAMOST spectral types versus Gaia spectral types of our sample of LELS.}
\label{fig:LAMOST_GAIA_spt}
\end{figure}

\subsubsection{Identifying PtMS ELS candidates based on log(\emph{g}) and CMD estimates} \label{sub_sec:giant_cut}

To segregate PtMS sources from our sample, we employed a series of three systematic steps, drawing upon the remaining sample of 42,891 sources. As a first step, we separated PtMS candidates using the available surface gravity (log($g$)) value from LAMOST. We segregated sources based on the criterion that the value of log($g$) exceeded three times its associated error ($\sigma$). Specifically, we applied the commonly used threshold of log($g$) $\leq$ 3.5, which is typically employed to distinguish between giants and dwarfs \citep{batalha2010selection,singh2019survey}. Applying the above criteria to the remaining sample, we classified 2,293 sources as PtMS candidates.

In addition, for M-type stars, the LAMOST pipeline classified its spectral types as dwarf M (dM)-type stars and giant M (gM)-type stars. In our remaining sample, 49 more sources belong to the gM type, which makes the new sample 2,342 PtMS candidates.


For the 7,618 sources lacking LAMOST log($g$) values, color-magnitude diagram (CMD) was employed to identify PtMS candidates within that sample. \cite{ruiz2018empirical} used Gaia-2MASS CMD to segregate Red Giant candidates in their sample. The availability of a larger sample and the presence of other predefined cuts to distinguish PtMS sources helped us retool the CMD cut to separate PtMS sources in our sample. For this purpose, we utilized Gaia CMD.

The criteria we defined to segregate our data is mentioned below:
\begin{align*}
    M_{G} &< 4.35 \\
    G_{BP}-G_{RP} &> 1 \\
    M_{G} &< 3.875 (G_{BP}-G_{RP}) - 1.3 
\end{align*}

The Gaia CMD of the LAMOST PtMS candidates is shown in Figure \ref{fig:giant_cut}. Applying the above mentioned criteria of magnitudes and colors we included an additional number of 2,327 sources to our sample. Combining all three samples obtained using the above criteria, we obtained a final sample of 4,669 PtMS candidates.



\begin{figure}
    \centering
    
\begin{tikzpicture}[node distance=2cm,thick,scale=0.7, every node/.style={scale=0.7}]

\tikzstyle{startstop} = [rectangle, rounded corners, minimum width=1cm, minimum height=1cm,text centered, draw=black]
\tikzstyle{io} = [trapezium,rounded corners, trapezium left angle=70,minimum height=0.8cm, trapezium right angle=110, text centered, draw=black]
\tikzstyle{process} = [rectangle,rounded corners, minimum width=2cm, minimum height=1cm, text centered, draw=black]
\tikzstyle{decision} = [diamond, rounded corners, text centered,aspect=2, draw=black,minimum width=2cm, minimum height=2cm]
\tikzstyle{arrow} = [thick,->,>=stealth]
\node (sub1) [startstop, text width=6cm] {LAMOST F0-M9 stars (8,613,834)};

\node (sub2) [decision,below of=sub1,yshift=-0.4cm,text width=3cm] {Shows emission in 6563 $\si{\angstrom}$ (80,860)};
\node (sub2_a) [right of=sub2, xshift=3cm] {Rejected};

\node (sub3) [io,below of=sub2,yshift=-0.1cm,text width=3.5cm] {SNR quality cuts (48,657)};
\node (pro_sub3) [right of=sub3, xshift=3cm] {Rejected};
\node (sub5) [io,below of=sub3,yshift=0.2cm,text width=5cm] {Photometric (2MASS, Gaia, and WISE) cross-match and its quality cuts (43,329)};
\node (pro_sub5) [right of=sub5, xshift=3cm] {Rejected};
\node (dec1) [decision,below of=sub5,yshift=-1.3cm, text width=4cm] {Selection method adopted from \cite{koenig2014classification} for class I and II sources};
\node (pro2a) [process, right of=dec1, xshift=5.5cm] {\textbf{Sources with IR excess} - (438)};
\node (sub6) [decision,below of=dec1,yshift=-3cm, text width=3.5cm] {log($g$) $<$ 3.5 OR \\ Spectral type: gMe OR\\ (M$_{G}$ $<$ 4.35  AND\\ G$_{BP}$-G$_{RP}$ $>$ 1 AND\\ M$_{G}$ $<$ 3.875 (G$_{BP}$-G$_{RP}$) - 1.3)};
\node (dec_sub6) [process, right of=sub6, xshift=5.6cm] {\textbf{Class PtMS*} - (4,669)};
\node (sub7) [decision,below of=sub6, yshift=-2.7cm, text width=3.5cm] {Presence of [NII] and S[II] emission lines};
\node (dec_sub7) [process, right of=sub7, xshift=4.5cm] {\textbf{Forbidden line sources} - (5,177)};
\node (dec1_sub8) [process, below of=sub7,yshift=-1.3cm, text width=5cm] {\textbf{Class Fe} if spectral type F* (1595) \\ \textbf{Class Ge} for spectral type G* (1,154) \\ \textbf{Class Ke} for spectral type K* (7,032) \\ \textbf{Class dMe} for spectral type dM* (23,264)};

\draw [arrow] (sub1) -- (sub2);
\draw [arrow] (sub2) -- node[anchor=east] {yes}   (sub3);
\draw [arrow] (sub2) -- node[anchor=south] {no}  (sub2_a);

\draw [arrow] (sub3) -- node[anchor=south] {no}   (pro_sub3);
\draw [arrow] (sub3) -- (sub5);
\draw [arrow] (sub5) -- node[anchor=south] {no}   (pro_sub5);
\draw [arrow] (sub5) -- (dec1);
\draw [arrow] (dec1) -- node[anchor=south] {yes} (pro2a);
\draw [arrow] (dec1) -- node[anchor=east] {no} (sub6);
\draw [arrow] (sub6) -- node[anchor=east] {no} (sub7);
\draw [arrow] (sub6) -- node[anchor=south] {yes} (dec_sub6);
\draw [arrow] (sub7) -- node[anchor=south] {yes} (dec_sub7);
\draw [arrow] (sub7) -- node[anchor=east] {no} (dec1_sub8);

\end{tikzpicture}
\caption{Flowchart of the classification criteria of LELS.}
\label{flowchart}
\end{figure}
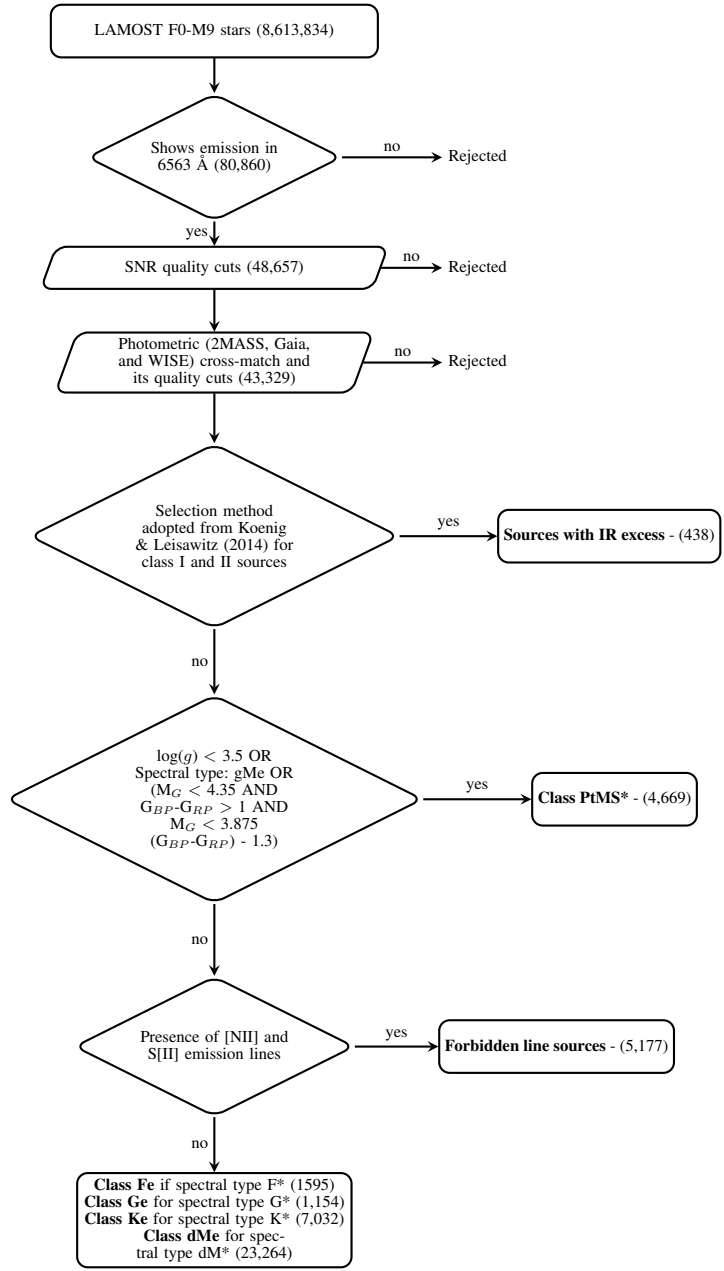

\subsubsection{Fe/Ge/Ke/dMe LELS candidates} \label{sub_sec:FGKM_cut}

From the rest of the sample (38,222 sources), sources exhibiting [NII] and [SII] in emission are separated and categorized as ``forbidden-line sources", totaling 5,177 sources. These stars will be further discussed in Shridharan et al. (under prep.). The remaining sample was classified into Fe/Ge/Ke stars and dMe stars based on their spectral types provided by LDR6. Since the LAMOST spectral type has been used as a classification scheme, validation of the spectral type for LELS sources is necessary. Hence, we plotted a comparison of the LAMOST spectral type with the Gaia spectral type of our sources, as shown in Figure \ref{fig:LAMOST_GAIA_spt}. The Gaia spectral types are estimated using General Stellar Parametriser (GSP) modules \citep{vallenari2023gaia}. The figure shows that other than G-type stars, all the other spectral types provided by LAMOST are fairly accurate.

Furthermore, the reason behind the separation of dMe stars from other LELS is the presence of intense molecular bands in M-type stars. None of these stars show IR excess. Hence, this sample contains a mix of potential class III sources, Active main-sequence LELS, and different types of variable stars. Further studies on young stars in this class are conducted by \cite{nidhi2023JApA...44...75N}. Figure \ref{flowchart} shows the flowchart of the entire classification scheme.

\section{Results} \label{sec:result}

In this section, we discuss the distribution of LELS in the Galaxy and the classification based on the major spectral features. 

\begin{table}
\centering
\begin{tabular}{||c | c||} 
 \hline
 Type & ELS Candidates \\ [0.5ex] 
 \hline\hline
Total late-type stars & 8,613,834       \\
Stars with $H\alpha$  & 80,860         \\
After 2MASS, Gaia, WISE crossmatch & 43,329 \\ 
Young IR excess Sources      & 438             \\
PtMS Candidates     &   4,669         \\
Forbidden line sources     &  5,177     \\
Fe/Ge/Ke Candidates      & 9,781      \\
dMe Candidates      & 23,264\\[1ex] 
 \hline
\end{tabular}
\caption{Statistics of different classes of LAMOST LELS in our study.}
\label{tab:stat}
\end{table}

\subsection{Distribution of LELS across the spectral range}

\begin{figure*}
\centering
  \includegraphics[width=1.0\linewidth]{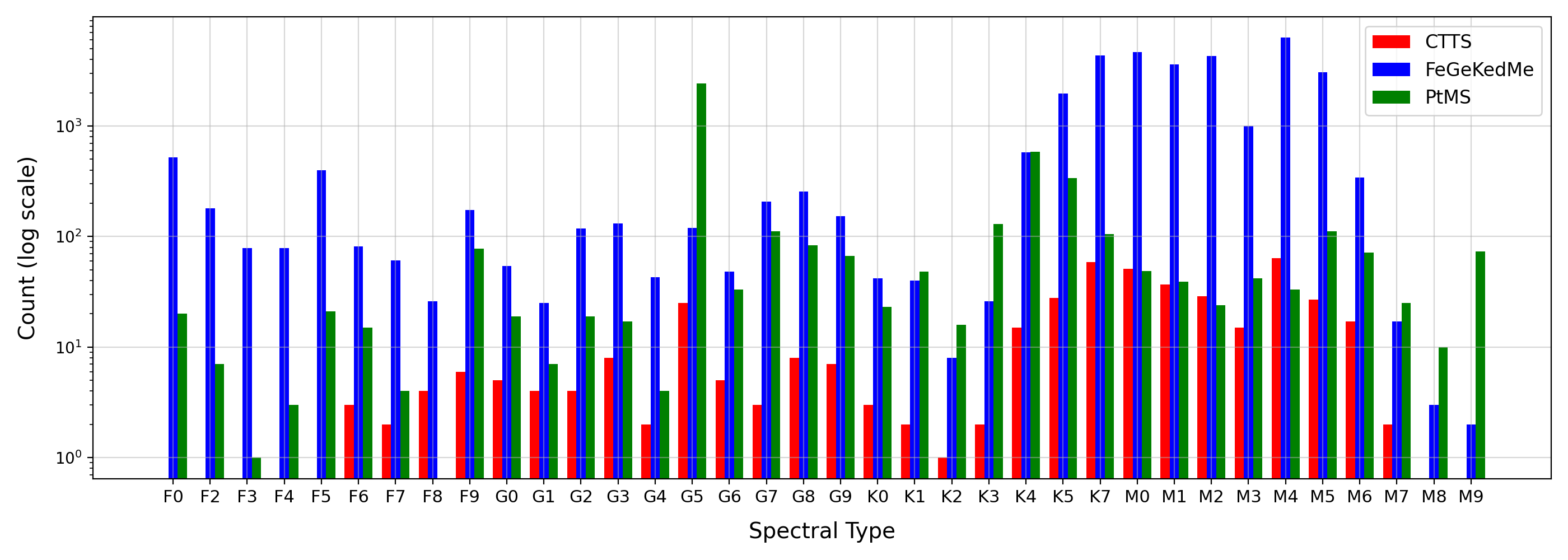}
  \caption{Histogram of LELS from LDR6. The spectra type is marked on the diagram along with different classes of LELS. The counts of stars belonging to each spectral type are represented in log scale.}
  \label{fig:num_distri}
\end{figure*}

Summarizing the classified sample, we have a dataset of 438 IR excess sources spanning across F6-M7 spectral types, 4,669 PtMS sources across F0-M9 spectral types, 9,781 Fe/Ge/Ke sources, and 23,264 dMe sources. This summary is tabulated in table \ref{tab:stat}. Furthermore, the distribution of LELS over the entire late-type spectral range is represented in Figure \ref{fig:num_distri}. From our sample of 38,152 LELS, we observed that the number of ELS increases between the spectral range of K5 to M5. When comparing with the total number of observed sources in LAMOST, this increase in number of ELS is very significant. This shows a direct correlation of higher activity in well-developed chromospheres. The number of stars steeply decreases after M6. The ELS that are observed after M5 are predominantly PtMS candidates. We conducted a crossmatch with SIMBAD to identify sources previously studied in the literature. This analysis revealed that 8,643 sources had already been documented, resulting in a newly detected sample of 29,509 LELS. A detailed description of the known sample of each class of LELS will be provided in section \ref{sec:Lit_Lels}.

\begin{figure}
\centering
  \includegraphics[width=1.0\linewidth]{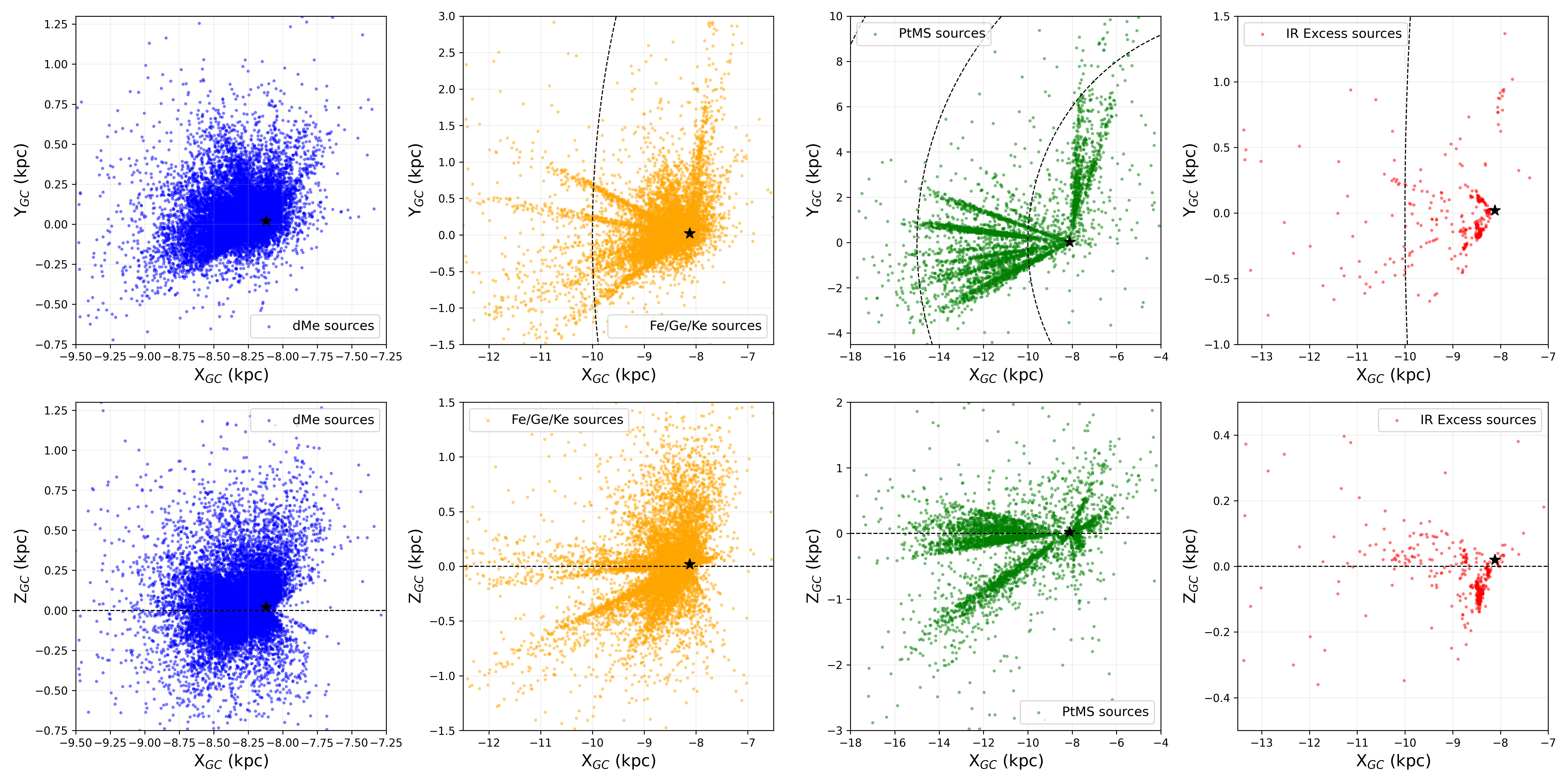}
  \caption{A Galactocentric distribution of LELS sources from LAMOST DR6. This coordinate system places center of our galaxy as the origin and (X$_{GC}$, Y$_{GC}$, and Z$_{GC}$) as the Cartesian coordinates. The Galactocentric X$_{GC}$ (in kpc) vs Galactocentric Y$_{GC}$ (in kpc) is plotted in the upper panel and Galactocentric X$_{GC}$ (in kpc) vs Galactocentric Z$_{GC}$ (in kpc) is plotted in the lower panel. Majority of the sources are observed in the anti-center direction. The LAMOST dMe stars (blue), PtMS candidates (green), Fe/Ge/Ke stars (orange), and IR excess sources (red) candidates are represented in this plot. The asterisk in all the plots shows the position of sun. The dotted line on the upper panel represents the galactic rings, representing distances of 10 kpc and 15 kpc from the galactic center. The dotted line on the lower panel represents galactic disk (Z$_{GC}$ = 0) plane.}
  \label{fig:polar}
\end{figure}

\subsection{3-D distribution of LELS in the Galaxy}\label{sec:data_dis}

Employing the methodology mentioned in Section \ref{subsec:classification}, we have compiled a sample of 38,152 LAMOST LELS. We have obtained the distances from \cite{bailer2021estimating}. Figure \ref{fig:polar} shows the galactocentric distribution of LELS present in our sample, where (-8000pc, 0pc) of (X,Y) represents the position of the sun. Around 81\% (30,145) of 38,152 LAMOST LELS are at a distance less than 1 kpc. Among them, 22,862 are dMe stars, which constitute over 98\% of the entire dMe sample spread over all the distances. More than 77\% (6,628) of the 9,781 Fe/Ge/Ke stars are observed within 1 kpc, and another 18\% (1,555) of them are observed between 1 and 2 kpc. In the case of IR excess sources, 320 out of 438 of these sources are observed within a distance of 1 kpc. The major contrast is observed in the case of PtMS candidates. Only 7\% (335) of 4,669 PtMS candidates are observed at a distance within 1 kpc. Over 72\% (3,343) is observed between 1 and 5 kpc. Similar to the spatial distribution observed in \cite{shridharan2021discovery}, the majority of LELS data is also distributed along the galactic anti-center direction.

\begin{figure}
\centering
  \includegraphics[width=1.0\linewidth]{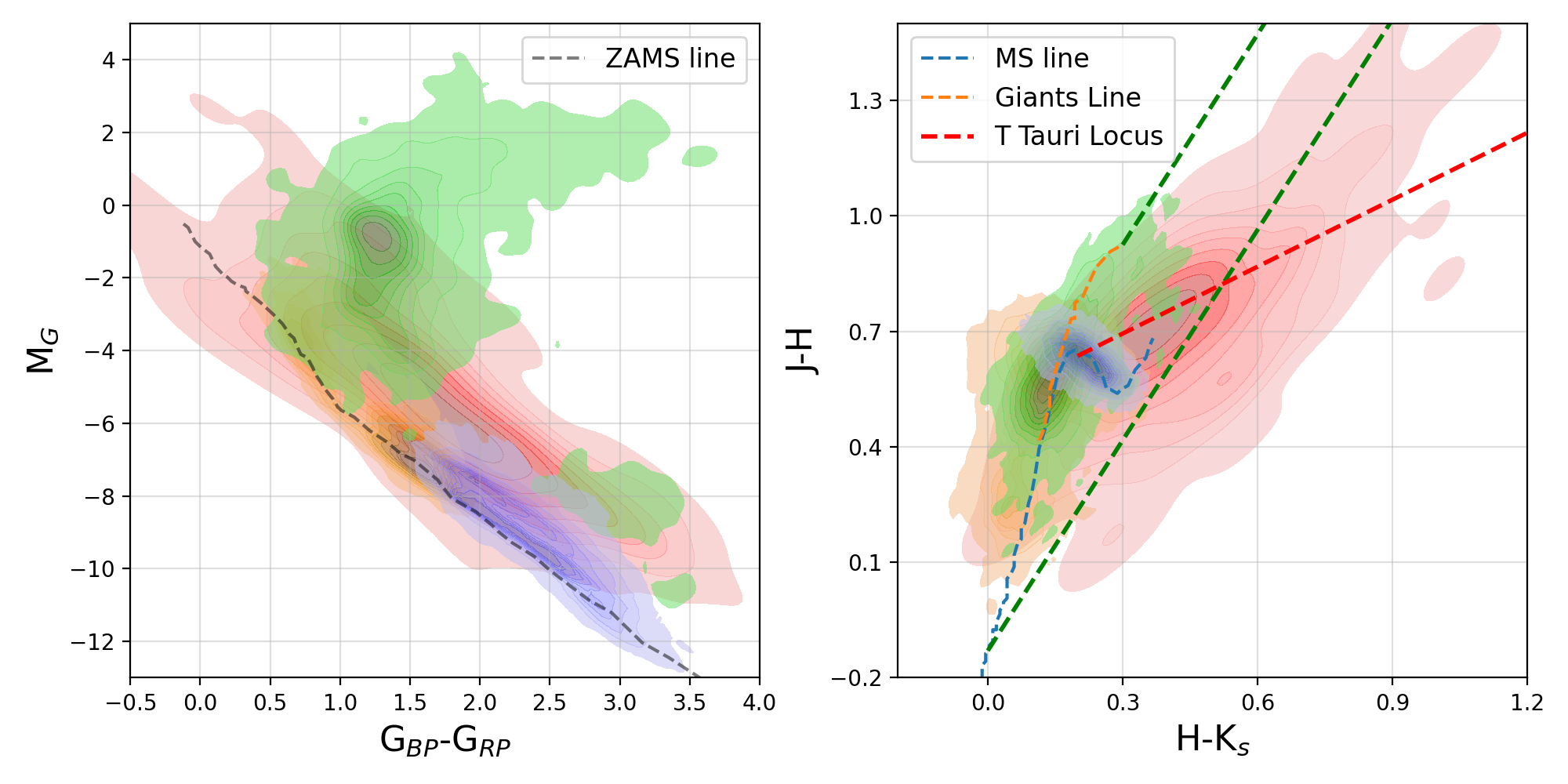}
  \caption{The KDE distribution of Gaia CMD (left) and 2MASS CCDm (right) of LAMOST LELS. The extinction correction of all the magnitudes has been done using the Bayester dust map \citep{2018JOSS....3..695M}. Both the plots represents dMe stars (blue distribution), Fe/Ge/Ke sources (yellow distribution), PtMS candidates (green distribution), and IR excess sources (red distribution). Also, ZAMS line (black dotted line), main-sequence loci (blue dotted line), Giants line (orange dotted line), and TTS locus (red dotted line) are also shown \citep{carpenter2001color}.}
  \label{fig:2mass_gaia_ccdcmd}
\end{figure}

\subsection{Photometry of LELS}

We utilized photometric data provided by 2MASS and Gaia to further study the distribution of our LELS dataset. Figure \ref{fig:2mass_gaia_ccdcmd} shows the kernel density estimate (KDE) photometric distribution of our LELS sample. It also helps us to validate the classification schemes utilized for our sample. CMD of our LELS sources are represented in the left panel of figure \ref{fig:2mass_gaia_ccdcmd}. The zero-age-main-sequence (ZAMS) line has also been shown in figure. The IR excess sources distribution (represented in red) is located in the region above the ZAMS line, whereas Fe/Ge/Ke sources (represented in yellow) and dMe sources (represented in blue) are over-plotted on the ZAMS line. The PtMS sources (represented in green) highlight the sub-giant, AGB, and RGB regions of the HR diagram \citep{eyer2019gaia}.

The CCDm of the LELS is shown in the right panel of Figure \ref{fig:2mass_gaia_ccdcmd}. We used the same color scheme for the CCDm as described for the CMD. It has been observed that majority of the IR excess sources are located around the T-Tauri locus \citep{meyer1997intrinsic}, as shown in the figure. We also noticed that the PtMS sources from our sample are clustered around the `giant locus' \citep{carpenter2001color}. The rest of the sample is clustered around the main sequence line. These distributions bring novelty to our classification scheme and bring validity to our sample.

\begin{figure}
\centering
  \includegraphics[width=1.0\linewidth]{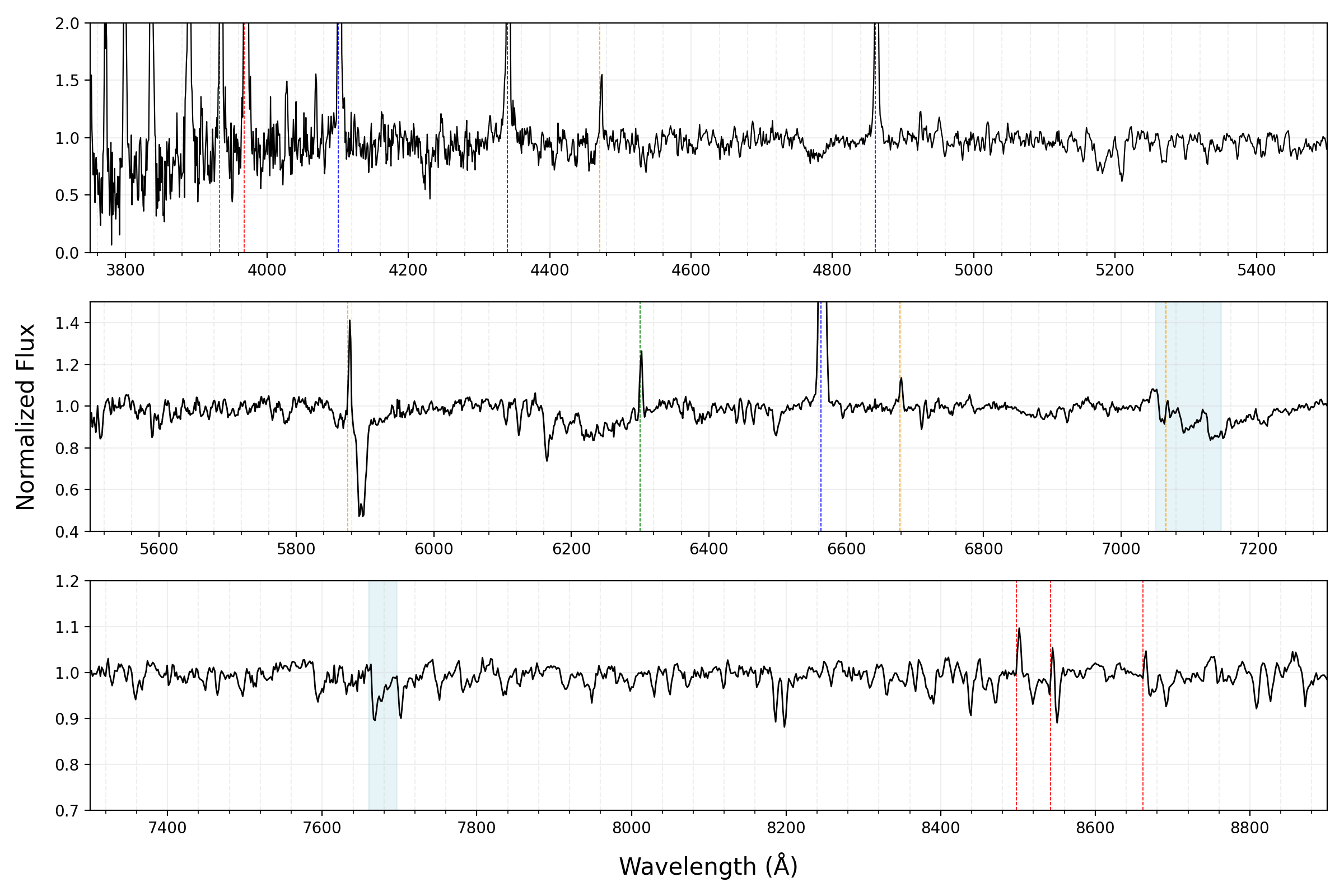}
  \caption{Visualization of the representative spectrum of a LAMOST T Tauri star. The Balmer lines are represented in blue, $Ca~{\textsc{ii}}$ lines in red, [OI] in green, and HeI lines in orange. Furthermore, TiO bands are highlighted in blue. The spectra belong to the star IQ Tau, a T Tauri star of spectral type M2. }
  \label{fig:tts_rep_spec}
\end{figure}

\subsection{Spectral features of LELS}

In this section, we mention the major emission lines observed in late-type stars. The major features include Balmer lines, metallic lines like $Ca~{\textsc{ii}}$ $H \& K$, CaIRT, $Mg$, and $Na$ doublet, and molecular bands such as $TiO$ and $VO$ for stars of spectral types later than K5. Figure \ref{fig:tts_rep_spec} shows the representative spectrum of a sample IR excess source. 

As we discussed in the previous sections, all the LAMOST LELS star spectra show $H\alpha$ in emission. The strength of this line is indicative of the circumstellar activity of the star or the stellar system. Other Balmer lines like $H\beta$, $H\gamma$, and $H\delta$ are also observed in a number of the LELS. The statistics of these observations, along with the line strength measurements are studied.

\begin{figure}
\centering
  \includegraphics[width=1.0\linewidth]{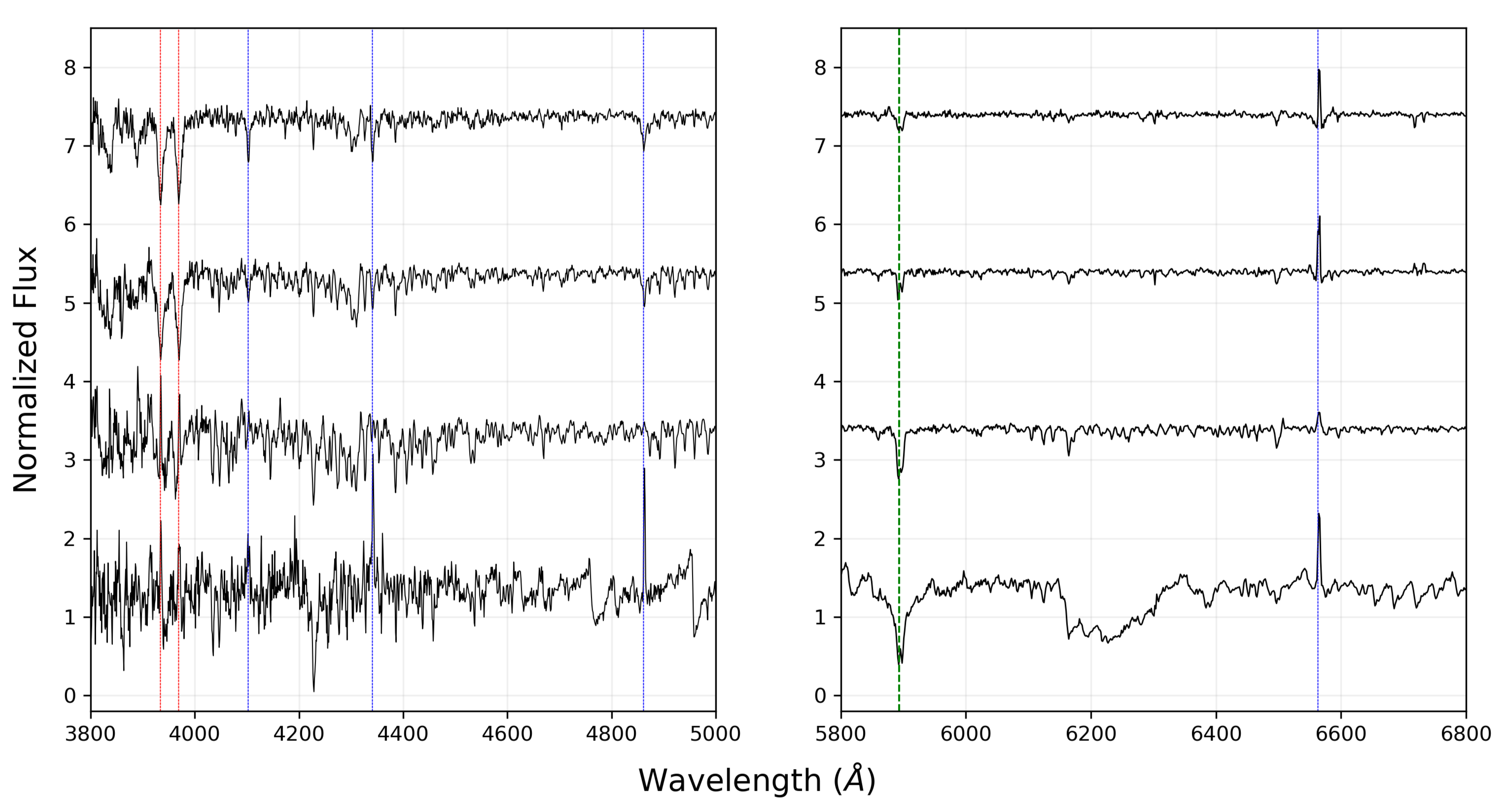}
  \caption{The representative spectra of a sample LAMOST Fe/Ge/Ke/Me candidate. The Balmer lines are represented in blue, the $Ca~{\textsc{ii}}$ H $\&$ K lines are represented in red, and NaI doublet in green. The spectral types of each of these individual spectra are also represented.}
  \label{fig:FGKM_spec}
\end{figure}

Other major emission lines observed include the $Ca~{\textsc{ii}}$ H and K lines, which are observed either in emission or absorption. These lines, along with Balmer lines, are presented in the literature as chromospheric activity tracers. Another significant ionized calcium line observed in LELS is the CaIRT ($\lambda$8498, 8542, 8662 $\si{\angstrom}$). Except for a number of IR excess sources, very few stars show triplet emission in their spectra.

Given their late-type classification, it is common to observe metallic lines in the spectra of these stars. The main metallic lines include $Na~{\textsc{i}}$ doublets at 5890, 5896 $\si{\angstrom}$, $Na~{\textsc{i}}$ ($\lambda$ 8183, 8195 $\si{\angstrom}$), $Mg~{\textsc{i}}$ ($\lambda$ 5184 $\si{\angstrom}$), $K~{\textsc{i}}$ ($\lambda$ 7665, 7701 $\si{\angstrom}$), $Fe~{\textsc{i}}$ ($\lambda$ 4383, 5270 $\si{\angstrom}$), $Ca~{\textsc{i}}$ ($\lambda$ 4226 $\si{\angstrom}$), etc. These lines are often observed in absorption. A few other metallic lines like $Fe~{\textsc{ii}}$ ($\lambda$ 5018, 5168 $\si{\angstrom}$), $Fe~{\textsc{i}}$ ($\lambda$ 4063, 5162 $\si{\angstrom}$), forbidden line of [$O~{\textsc{i}}$] ($\lambda$ 6300 $\si{\angstrom}$) are also observed in emission. Figure \ref{fig:FGKM_spec} displays representative spectra from the Fe, Ge, Ke, and dMe classes. 

The major molecular bands observed in late-type stars include $TiO$ (band heads at $\lambda$ 4760, 5160, 7050, and 7580 $\si{\angstrom}$), which are prime indicators of the temperature of a star. These bands dominate the spectra of late-K to M-type stars. Other major molecular bands observed include $VO$ (band heads at $\lambda$ 7010, 7383 $\si{\angstrom}$) and $CaH$ (band heads at $\lambda$ 6400, 6800 $\si{\angstrom}$).

The emission line profile of our sources shows various types of line features. The intensity of emission varies depending on the level of activity in these sources. The mass column density of the chromospheres and temperature structures inside the chromosphere will significantly affect the line profiles and line intensity of low mass sources \citep{houdebine1995observation}. The literature shows that most of the profiles can be attributed to the inclination angle, stellar rotation, or chromospheric activity of these sources. Furthermore, strong asymmetry in emission profiles have been observed in red giants and LPVs \citep{gillet1988balmer}. Given the large sample size of nearly 40,000 LELS, classifying the line profile of each source presents a significant challenge and falls outside the scope of this catalog. This detailed classification will be reserved for future studies, where individual groups of LELS will be analyzed in greater depth \citep{Anusha2021MNRAS.501.5927A,nidhi2023JApA...44...75N}. 

In our work, we have estimated the equivalent width of major lines like Balmer lines ($H\alpha$, $H\beta$, $H\gamma$, and $H\delta$), $Ca~{\textsc{ii}}$ H $\&$ K, and CaIRT using the \texttt{line\_index} module of the \texttt{laspec} package using an automated Python routine. This module gives us the values of equivalent width and equivalent width error for all the respective lines. Within our study, a value for the emission-line equivalent width is deemed legitimate only if the matching equivalent width value is three times the projected error ($\sigma$) supplied by $laspec$ ($EW>3 \sigma$).


\subsubsection{Balmer lines}

$H\alpha$ in emission is one of the major indicators of activity in low-mass stars. In sources with IR excess, predominantly in young stars, the emission arises from the magnetospheric columns \citep{hartmann1994magnetospheric}, giving rise to very intense emission lines. From the literature, a significant correlation between stellar rotation and magnetic activity has been observed in low mass stars. Studies that have compared star rotation to activity characteristics have found that the relationship is valid for sources that are fully or partially convective \citep{delfosse1998rotation,reiners2012catalog}. Rotation rates will also play a major factor. A clear association has also been found between stellar rotation and EW($H\alpha$) \citep{newton2017halpha}.

\begin{figure}

\centering
  \includegraphics[width=1.0\linewidth]{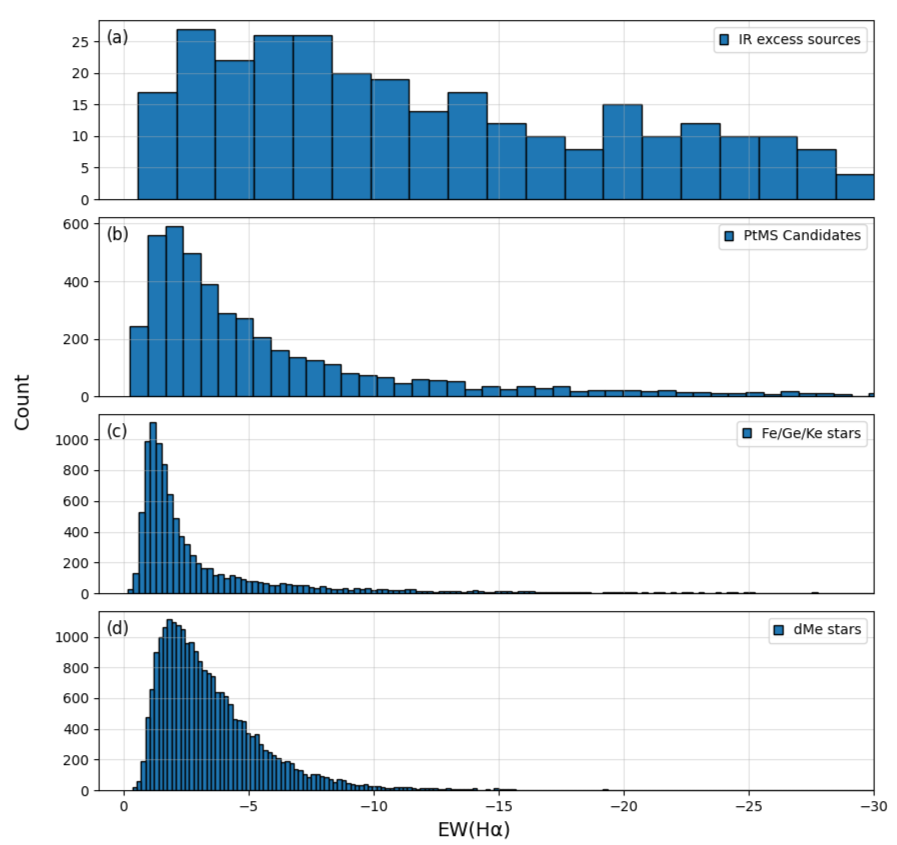}
  {\phantomsubcaption\label{fig:hist_ha_IR}%
     \phantomsubcaption\label{fig:hist_ha_gia}%
     \phantomsubcaption\label{fig:hist_ha_fgk}%
     \phantomsubcaption\label{fig:hist_ha_dMe}}%
  \caption{The figure represents the histogram of equivalent widths of $H\alpha$ of various classes of LELS. The histogram contains IR excess sources, PtMS candidates, Fe/Ge/Ke stars, and dMe candidates. }
  \label{fig:EW_hist_Ha}
\end{figure}

Figure \ref{fig:EW_hist_Ha} represents the histogram of equivalent width of $H\alpha$ in various classes of LELS. The bin size for the histogram is calculated using Freedman-Draconis rule (FDR) for each sample. For those with bin sizes estimated to be too small using FDR method, the bin sizes were provided manually. This is the rule followed throughout this work for histogram distributions. 

Figure \ref{fig:hist_ha_IR} represents the histogram distribution in the $H\alpha$ strength (EW($H\alpha$)) for IR excess sources. Observations indicate that IR excess sources exhibit the most intense Balmer emissions compared to other classes. It has a median value of -19 $\si{\angstrom}$, as the strength is primarily due to disk accretion, as mentioned earlier. Figure \ref{fig:hist_ha_gia} shows the emission histogram distribution for PtMS sources, and it shows a median value of -4 $\si{\angstrom}$. The strength of the line is minimum for dMe candidates (fig \ref{fig:hist_ha_dMe}) with a median value of -3.2 $\si{\angstrom}$. The models of $H\alpha$ emission in TTS suggest that the emission strength can vary due to rotation, turbulence, and non-axisymmetric accretion in these sources \citep{kurosawa2005formation,wilson2022hydrogen}. Stellar winds are also one of the causes, which account for the increased line strength in CTTS \citep{lima2010modeling}. In the case of main-sequence late-type stars, the presence of $H\alpha$ emission is attributed to chromospheric and coronal activity. In solar-type sources, all active phenomena are directly correlated with the presence of intense magnetic fields generated by a dynamo mechanism \citep{charbonneau2014solar}. 
 

A total of 27,642 LELS show emission in $H\beta$. It includes 341 IR excess sources, 2,779 PtMS candidates, 3,703 Fe/Ge/Ke stars, and 20,819 dMe candidates. Over 86$\%$ of IR excess sources and more than 85$\%$ of dMe candidates show emission in $H\beta$. Only 35$\%$ of PtMS candidates and 50$\%$ of Fe/Ge/Ke candidates show $H\beta$ emission. The strength of emission ranges from -0.84 to -100 $\si{\angstrom}$, with a median value of -2 $\si{\angstrom}$. 


\subsubsection{$Ca~{\textsc{ii}}$ emission lines}

$Ca~{\textsc{ii}}$ emission lines are observed in late-type stars, both in the bluer and redder ends of the spectra. We observed $Ca~{\textsc{ii}}$ H \& K lines at wavelengths 3968 $\si{\angstrom}$ and 3933 $\si{\angstrom}$, respectively and the CaIRT are observed at wavelengths 8498, 8542, and 8662 $\si{\angstrom}$. \cite{mould1977band} studied these emission lines and their dependencies with luminosity and precisions on other physical attributes in these objects. 

From our sample, it has been observed that 4,887 LELS show emission in $Ca~{\textsc{ii}}$ H, among which 164 (37$\%$) are IR excess sources, 141 (3$\%$) are PtMS candidates, 722 (8$\%$) are Fe/Ge/Ke stars, and 3,860 (17$\%$) are dMe stars. The presence of emission/absorption contribution from the HI line in $Ca~{\textsc{ii}}$ H suggests the necessity of distinguishing between the contributions of both for subsequent analyses. This separation is crucial for future investigations. 

\begin{figure*}

\centering
  \includegraphics[width=1.0\linewidth]{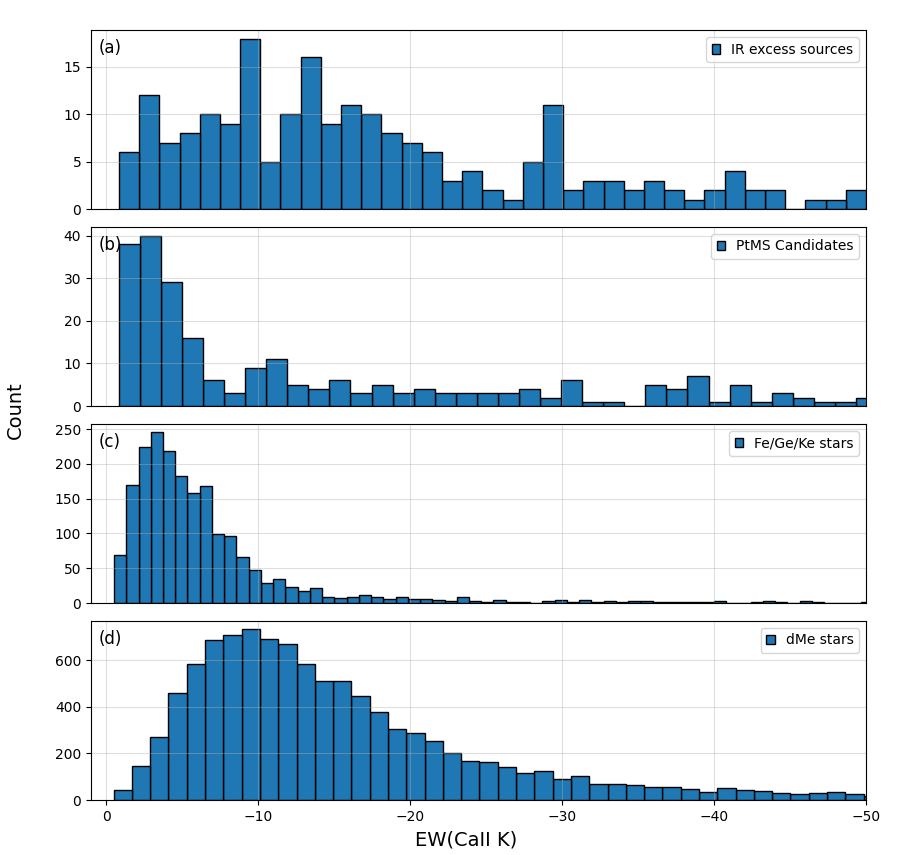}
  {\phantomsubcaption\label{fig:Hist_CaIIK-1}%
     \phantomsubcaption\label{fig:Hist_CaIIK-2}%
     \phantomsubcaption\label{fig:Hist_CaIIK-3}}%
  \caption{The distribution of equivalent widths of major emission lines over the entire range of late spectral type of LAMOST LELS is shown in the figure. Fe/Ge/Ke/dMe sources, PtMS candidates, and IR excess sources are shown in the plot.}
\label{fig:Hist_CaIIK}
\end{figure*}

6,215 LELS show emission in $Ca~{\textsc{ii}}$ K, with equivalent width values in the range -2.5 to - 55 $\si{\angstrom}$. Around 39$\%$ of IR excess sources (170), more than 18$\%$ of dMe stars (4,315), over 3$\%$ of PtMS sources (155), and close to 18$\%$ of Fe/Ge/Ke sources (1,575) show emission in the $Ca~{\textsc{ii}}$ K line. This emission line is prominently observed in M-type stars, i.e., stars having effective temperatures of $T_{eff} < 4000K$. This holds in the cases of young IR excess sources as well. Figure \ref{fig:Hist_CaIIK} shows the histogram distribution of EW ($Ca~{\textsc{ii}}$ K) of various classes of LELS.

From the estimation of line parameters, it has been observed that the CaIRT in emission is seen rarely in main-sequence late-type stars. 893 stars show emission in the $Ca~{\textsc{ii}}$ 8498 $\si{\angstrom}$. More than 50\% of the IR excess sources (212), 505 dMe stars, and 86 PtMS candidates show emission in $Ca~{\textsc{ii}}$ 8498 $\si{\angstrom}$. The range of emission of $Ca~{\textsc{ii}}$ 8498 is from -0.4 to -17 $\si{\angstrom}$. The other two lines of the CaIRT show a similar equivalent width distribution. $Ca~{\textsc{ii}}$ 8542 emission is seen in 310 stars, of which 176 are IR excess sources, 62 are dMe stars, and 37 are PtMS candidates. $Ca~{\textsc{ii}}$ 8498 $\si{\angstrom}$ emission line, over 40$\%$ of IR excess sources show emission in this line, with equivalent widths ranging from -0.4 to -58 $\si{\angstrom}$. 323 stars show emission in the $Ca~{\textsc{ii}}$ 8662 $\si{\angstrom}$ emission. 178 IR excess sources and 50 PtMS candidates show emission in $Ca~{\textsc{ii}}$ 8662 $\si{\angstrom}$. 232 of the 38,152 LELS show emission in all lines of the CaIRT. 166 IR excess sources, 12 Fe/Ge/Ke stars, 28 dMe stars, and 26 PtMS candidates show emissions in all three lines.

The majority of the stars that show emission in these sets are IR excess sources. This confirms that the presence of CaIRT emission indicates the youth of these objects as well. Most of the IR excess sources belong to the class of T Tauri stars and the stellar properties and mass accretion rates of these stars are studied in \cite{nidhi2023JApA...44...75N}. 


\subsubsection{Paschen lines}

Paschen lines were also seen in the emission of some LELS. The equivalent widths of P12, P14, and P17 are calculated initially since these lines do not overlap with the $Ca~{\textsc{ii}}$ triplet. 916 LELS showed emissions in P12. The emission is seen in the range of -0.2 to -1.25. The line is mostly seen in Fe/Ge/Ke stars, as 422 of them show emission in P12, which includes 16 IR excess sources, 180 PtMS candidates, and 253 dMe stars show emission. The P14 line is seen in emission in 641 LELS. 242 Fe/Ge/Ke stars, 236 dMe stars, 98 PtMS candidates, and 12 IR excess sources show emissions ranging from -0.2 to -1.4. 1762 stars show emissions in the P17 line. 948 dMe stars, 597 Fe/Ge/Ke stars, 210 PtMS candidates, and 17 IR excess sources show P17 emission. Paschen emission is not common in IR excess sources, whereas $Ca~{\textsc{ii}}$ triplet is seen more often.


\subsection{Relation between Spectral type and major emission lines}

\begin{figure}

\centering
  \includegraphics[width=1.0\linewidth]{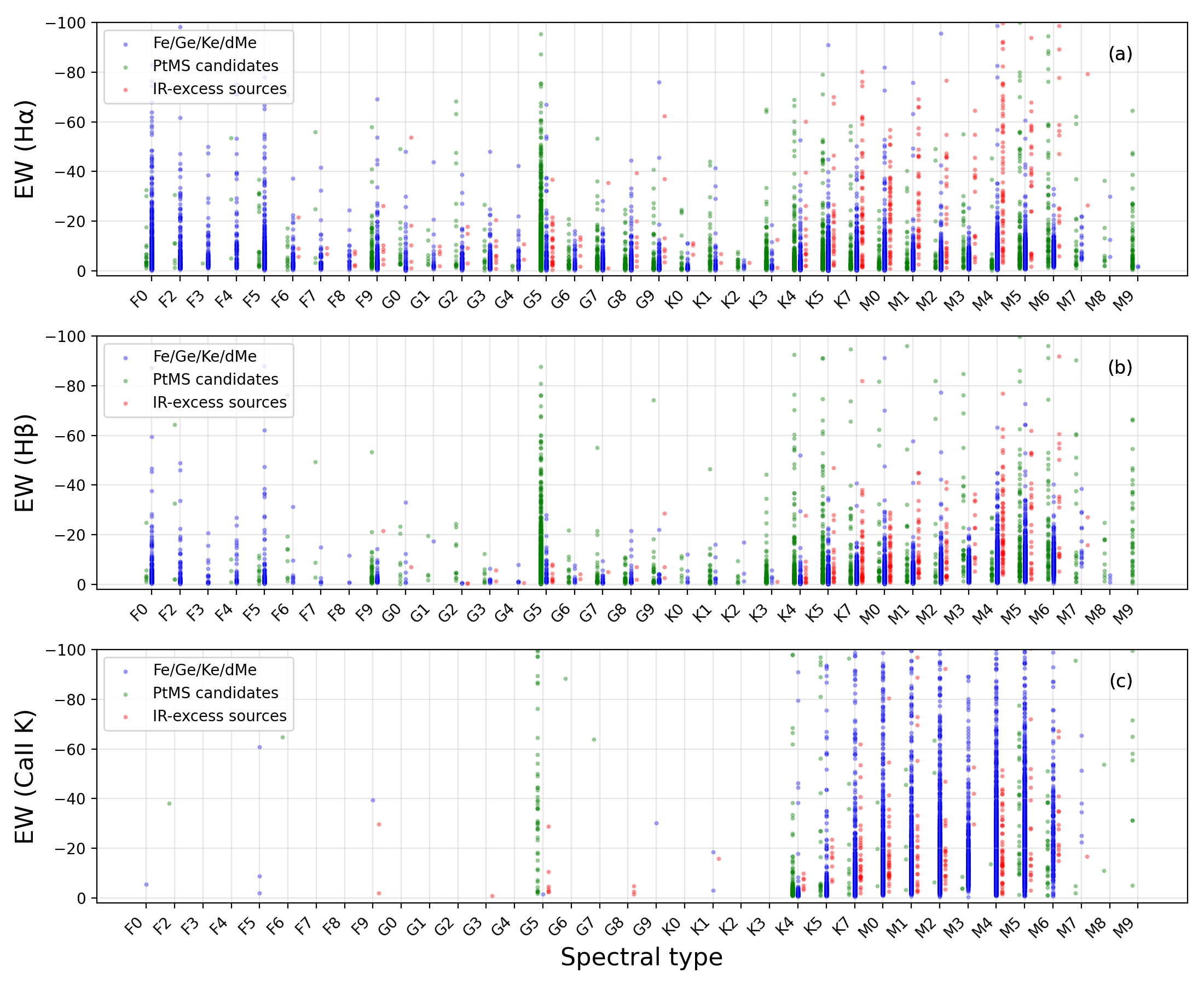}
  {\phantomsubcaption\label{fig:EW_spt-1}%
     \phantomsubcaption\label{fig:EW_spt-2}%
     \phantomsubcaption\label{fig:EW_spt-3}}%
  \caption{The distribution of equivalent widths of major emission lines over the entire range of late spectral type of LAMOST LELS is shown in the figure. Fe/Ge/Ke/dMe sources(blue), PtMS candidates (green), and IR excess sources (red) are shown in the plot.}
  \label{fig:EW_spt}
\end{figure}

Figure \ref{fig:EW_spt-1} shows the variation of the $H\alpha$ equivalent width with respect to the spectral type for IR excess sources, PtMS candidates, and Fe/Ge/Ke/dMe stars. Also, $H\beta$ shows a similar distribution over the late-type spectral range in emission as that of $H\alpha$, as shown in Figure \ref{fig:EW_spt-2}. The strength of emission from these sources peaks towards the later K- to M-type sources. Studying this allows for a deeper understanding of the emission trends across the spectral range and facilitates the comparison of the presence or absence of various emission lines. This, in turn, aids in elucidating the emission mechanisms in these sources.


$Ca~{\textsc{ii}}$ K is one of the most active chromospheric indicators \citep{hall2008stellar}. Figure \ref{fig:EW_spt-3} shows the emission strength of $Ca~{\textsc{ii}}$ K over various spectral types. The notable thing is the presence of emission of $Ca~{\textsc{ii}}$ K in LELS sources of spectral types later than K5. This is because of the presence of chromospheric activity in those sources. Among the stars in the spectral range F6-K4, less than 1\% of them show emission in $Ca~{\textsc{ii}}$ K, while more than 40 \% of sources later than K4 shows emission. This suggests the presence of active chromospheres and coronae in sources with spectral types later than K5 \citep{young1977study,hall2008stellar}.

\subsection{Literature crossmatch of known LELS}\label{sec:Lit_Lels}

The SIMBAD cross-match of these objects provided us with a sample of 373 stars from a total of 438 IR excess sources. Of the 373 known stars, 317 belong to four main classes. They are TTau*/TTau candidates, YSOs/YSO candidates, Orion V*, and Em* classes. Also, the classes ``Star" and V* contribute to 24 more candidates. These contribute to 94$\%$ of the known sample. Our sample has 65 newly detected IR excess sources.


The evolved stars are separated from other classes based on the method mentioned in Section \ref{sub_sec:giant_cut}. It provided us with a sample of 4,669 PtMS candidates. The SIMBAD crossmatch has provided us with a known sample of 944 stars. From that cross-matched sample, we observe that 160 stars belong to different PtMS star classes, which are asymptotic-giant branch stars (AGB*), horizontal branch stars (HB*)/HB* candidates, Mira variables/Mira candidates, red-giant branch stars (RGB*)/RGB candidates, red super-giants, and post-AGB*. Also, there are 88 Long Period Variable Stars (LPV*)/LP* candidates. 81 of the known samples belong to binary classes like eclipsing binary stars (EB*) or EB* candidates, symbiotic stars (Sym*), spectroscopic binaries (SB*), and RS Canum Venaticorum (RS CVn). Variable star classes like pulsating variables (Pul V*), eruptive variables (Eruptive*), variable stars (V*), ELS (Em*), Orion variables (Orion V*), BY Draconis variables (BY Dra), and pulsating variables (Pul V*) contribute to 150 more known stars. 115 stars are classified as YSOs as well. The presence of YSOs in this sample is evident from their location in the HR diagram and the lack of IR excess. These might be candidates of Weak-line T-Tauri stars, which show an absence of IR excess and weak $H\alpha$ emission \citep{alcala1993t}. Also, 293 stars were represented in the `Star' class, and 14 were represented as low-mass stars (low-mass*). A couple of stars belong to groups such as cepheids and far-infrared (FIR) sources, which sums up the known sample of PtMS stars. We have a sample of 3,725 new detections, which doesnt have any information on SIMBAD.

A SIMBAD cross-match of the dMe stars revealed that 5,731 are previously detected stars, while 17,533 sources are new detections. Of the known sample, 1,490 are categorized as `star/**' in SIMBAD. 1,032 were mentioned as stars with higher proper motion (PM*). 1083 stars belong to various variable star groups like V*, Em*, Orion V*, and Eruptive*. 301 belong to binary groups such as EB*, SB*, and RSCVns. This group also has 1,322 YSO/YSO candidates. Other classes such as Pec*, mid-infrared (MIR), near-infrared (NIR), and radio sources also have a couple of stars each in every group.

From our sample of 9,781 Fe/Ge/Ke stars, 1,595 are Fe-type, 1,154 are Ge-type, and 7,210 are Ke-type. After the SIMBAD crossmatch of this sample, 1,882 stars are determined as the known sample. Like other classifications, most of our known sample belongs to some variable star classes like Orion V*, V*, Em* (ELS), Eruptive*, PulsV*, and RotV*, which contribute to 484 stars. 531 of them are just represented as ``stars", which can be converted to ELS after our LAMOST spectroscopic results. 281 are classified as young stars belonging to classes YSO/YSO candidates and Ttau*/Ttau* candidates. 315 of them are classified into binary star groups like EB*/EB* candidates and SB*. We have a sample of 7,899 Fe/Ge/Ke sources as new detections. Our sample of Fe/Ge/Ke stars can be suitable candidates for further variability study to enrich the understanding of various active phenomena in these sources.

\subsection{Classification of LELS based on the presence of emission lines}\label{groups}

As mentioned in the previous sections, $H\alpha$, $Ca~{\textsc{ii}}$ H \& K, and Ca IRT are the most commonly observed emission lines in LELS. The presence or the absence of more than one of these emission lines will be of great physical significance. Hence, for our sample, we devised a classification scheme of LELS based on presence or absence of three major emission lines, namely $H\alpha$, $Ca~{\textsc{ii}}$ K (as a representative of $Ca~{\textsc{ii}}$ H \& K), and $Ca~{\textsc{ii}}$ $\lambda$ 8542 $\si{\angstrom}$(the representative line from the $Ca~{\textsc{ii}}$ triplet). Since we are using the entire range of spectra for this study, we had to consider some initial assumptions in terms of SNR at various spectral regions, especially in the \emph{u} band. After visual inspection of the spectra, SNR$_{u}$ \textgreater{} 0.8 was given as the selection criteria for this study.

\begin{figure}
\centering
  \includegraphics[width=1.0\linewidth]{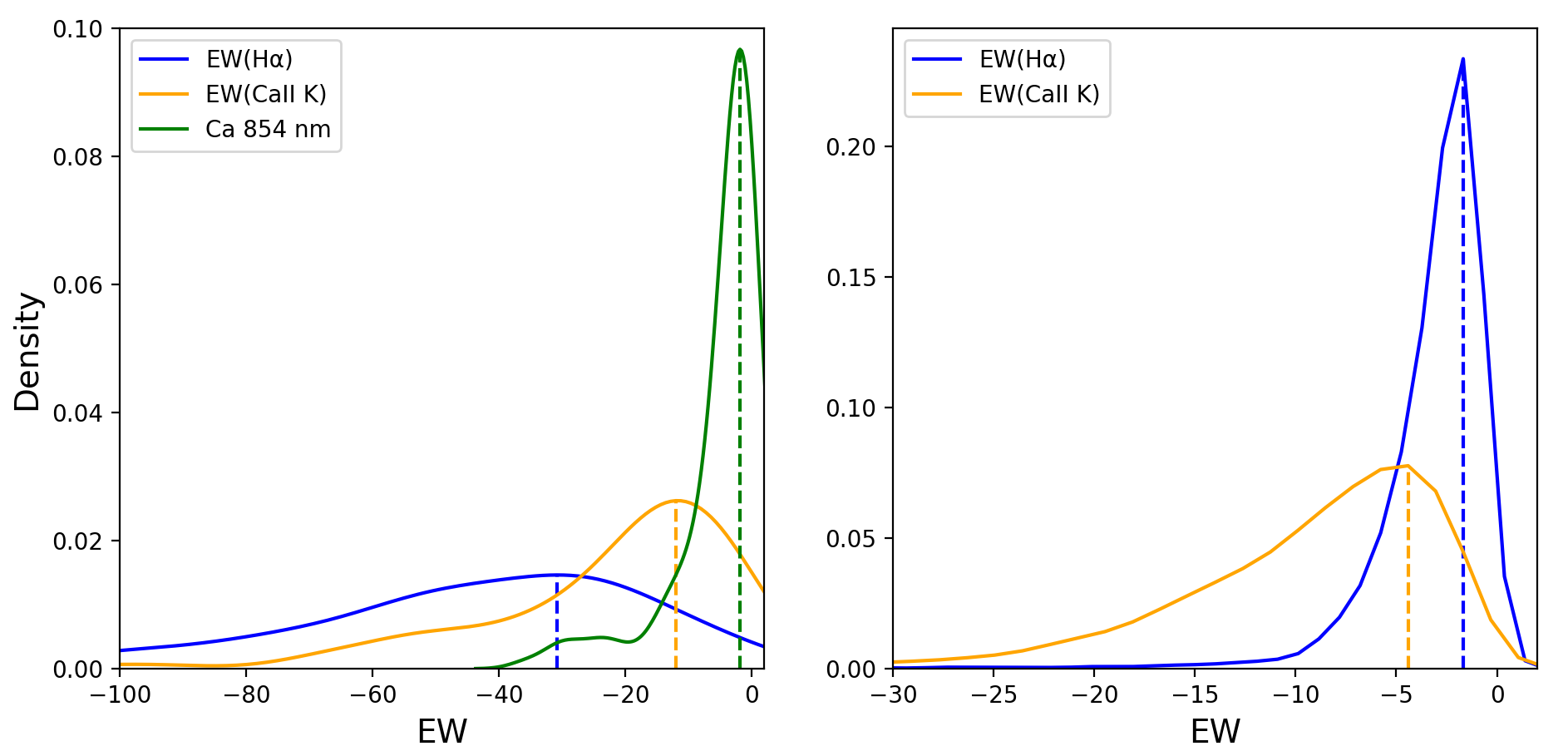}
  {\phantomsubcaption\label{fig:hist_grp_1}%
     \phantomsubcaption\label{fig:hist_grp_2}}%
\caption{The KDE distribution of equivalent widths of emission lines from Group A (left) and group B (right). $H\alpha$ is represented in blue, $Ca~{\textsc{ii}}$ K in orange and Ca II 8542 $\si{\angstrom}$ in green. The corresponding dotted lines represents most number of stars showing same equivalent width.}
\end{figure}

\begin{itemize}
    \item Group A comprises stars displaying emission in $H\alpha$, Ca II K, and Ca IRT, totaling 142 stars. Intense emission lines are observed in this group, with a median $H\alpha$ equivalent width of -41 $\si{\angstrom}$. More than 71\% (101) of stars in this group belong to IR excess sources, 13.4\% (19) belong to PtMS candidates, and 14\% (20) belong to dMe sources. It was also noticed by \cite{muzerolle1998emission} that HeI emission is also observed in these sources. We have observed that 93 sources show emission in HeI $\lambda$ 5876 $\si{\angstrom}$ in these sources. It is confirmed by the presence of weaker HeI $\lambda\lambda$ 6678, 7605 $\si{\angstrom}$ emission lines. Moreover, the literature crossmatch revealed 66 stars classified as YSO/TTau stars and 47 categorized as Em*/OrionV* stars, indicating majority of the sources belong to disk-like systems. This confirms the presence of young and active sources within this group. Figure \ref{fig:hist_grp_1} shows the KDE equivalent width distribution of $H\alpha$, Ca II K, and Ca IRT emission lines. This set is significant due to its implication on the line origin of emission of all three lines, indicating disk emission.  
    
    \item Stars that show emission in $H\alpha$ and $Ca~{\textsc{ii}}$ K are classified as Group B. This group contains 5,898 stars, which include 66 IR excess sources, 133 PtMS candidates, 1,548 Fe/Ge/Ke stars, and 4,151 dMe stars. A noteworthy observation is that only 10 stars, spanning all classes, have a spectral type less than K4. Also, we observed that 94\% (5,542) of sources exhibit a greater emission strength of Ca II K compared to $H\alpha$. Among the 2469 known stars within this group, 63\% (1,552) are classified under various variable star categories, including eruptive variables (Eruptive*), Orion variables, BY Draconis variables, etc., indicating majority of sources showing rotation coupled with starspots and other chromospheric activities. Figure \ref{fig:hist_grp_2} shows the KDE distribution of the above mentioned lines. The strength of lines implicate that the lines originate primarily from the chromospheres in these sources. The strength difference in both these emission lines as mentioned above, also indicates the same scenario. 
    
    \item Group C contains stars that shows emission in $H\alpha$ and CaIRT in emission. 25 stars belong to this group. It has been observed that the majority of the sample, along with the CaIRT, show Paschen lines in emission. Of the 25 stars, 9 are associated with IR excess sources, 8 with Fe/Ge/Ke sources, 2 with PtMS sources, and 6 with dMe sources. Since the number of sources in group C is quite low, it suggests the probability of both $H\alpha$ and Ca IRT being in emission is minimal. 
    
    \item Group D contains stars that shows emission in $H\alpha$ alone. We have a sample of 3,785 stars in this group. It contains 49 IR excess sources, 1,164 PtMS candidates, 2,057 Fe/Ge/Ke stars, and 515 dMe stars. A SIMBAD crossmatch of the known sample identified 1,534 sources already studied in the literature. This group includes post-main-sequence (PtMS) stars such as AGB*, RGB*, and Mira variables, as well as other variable star categories like Em*, V*, and OrionV*. This diversity indicates a wide range of activity among these sources, including mass loss, rotation, and chromospheric activity. 
\end{itemize}

To conclude, majority of sources in Groups A belong to the IR excess sources class. The line strength of major emission lines is also very intense for those sources. In Group B, the emission is predominantly due to chromospheric activity, suggested by the presence of a large number of sources belonging to Fe/Ge/Ke/dMe sources. It is also interesting to note the lack of sources earlier than K4 in this group, suggesting intense chromospheric activity in sources later than K4.

\subsection{Catalog description}
The catalog of LAMOST LELS containing 38,152 sources, which include 438 infrared excess sources, 4,669 post-main-sequence candidates, 9,718 Fe/Ge/Ke sources, and 23,264 dMe sources. We further classified 9850 sources into four groups based on the presence or absence of Balmer and Ca emission lines, as shown in section \ref{groups}. The sources are named with a common prefix of LM-LEMC (Low-Mass LAMOST EMmision-lne Catalog) followed by the index. we have sources ranging from LM-LEMC\_1 to LM-LEMC\_38152. A sample table is shown in Table \ref{table:Catalog}. The data include the LAMOST design ID, positions, our classification and grouping. The catalog developed for our LELS sample will be made available online. The online table will include the photometric values and the estimated line equivalent widths.

\section{Conclusion}


In this study, we present the spectroscopic information of 38,152 LELS, identified from the LAMOST DR6 data release. Acquiring F0-M9 ``STAR" data from LAMOST DR6, we utilized a custom Python routine developed with the $\texttt{scipy.signal.find peaks}$
parameter to isolate spectra containing $H\alpha$ emission. As the next step, we separated the ELS into various sub-classes such as PtMS candidates, IR excess sources, Fe/Ge/Ke stars, and dMe stars. We employed photometric and spectroscopic methods to do the same.

We have obtained one of the largest homogeneous datasets of LELS in this study. After the SIMBAD cross-match, a sample of 29,222 was found to be new detections. This includes 65 IR excess sources, 7,899 Fe/Ge/Ke stars, 17,533 dMe stars, and 3,725 PtMS candidates. The spatial distribution of our data suggested that the stars observed are in the anti-center direction, which is similar to what we observe in the early-type ELS work done by \cite{shridharan2021discovery}. Also, over 89\% of the stars observed were within 1 kpc. It has also been observed that PtMS star candidates are observed at much longer distances due to their size and high luminosity.



We also discussed the spectral features observed in various late-type stars, encompassing both absorption and emission lines. Studying the major spectral lines observed in a star helps us understand the chemical composition, active phenomena, and the physical properties of a star. For late-type stars, the major emission features observed in our sample are noted. Furthermore, the line strength of the major emission lines, which include Balmer lines, ionized Ca lines, Pa lines, etc., and its distribution over various spectral types were analyzed. The statistics revealed that we observe intense emission lines in young stars due to their youth and intense disk activity in these sources. We observed strong metallic absorption lines in the spectra of F-, G-, and K-type ELS. Further, we have classified the stars into 4 groups based on the presence or absence of 3 major emission lines, namely $H\alpha$, $Ca~{\textsc{ii}}$ K, and and $Ca~{\textsc{ii}}$ 8662 $\si{\angstrom}$. More deeper studies in this direction will help us to understand the line formation scenarios in low-mass stars.

The distribution of stars over various spectral types is also discussed in this work. Even though the classifications are made for various groups of LELS, further studies on various groups are required for a better understanding of the various properties of these objects. 

\section*{Acknowledgement}

We express our gratitude to the reviewer for their valuable comments and suggestions, which have greatly enhanced the quality of this manuscript. We would like to thank the Center for Research, CHRIST (Deemed to be University), Bangalore, India for providing the necessary support. B.M. is grateful to the Centre for Research, CHRIST (Deemed to be University), Bangalore for the research grant (Seed Money project) to carry out the present project (SMSS-2335). This work has made use of data products from the Guo Shoujing Telescope (the Large Sky Area Multi-Object Fibre Spectroscopic Telescope, LAMOST), and data from the European Space Agency (ESA) mission Gaia (\url{https://www.cosmos.esa.int/gaia}), processed by the Gaia Data Processing and Analysis Consortium (DPAC, \url{https://www.cosmos.esa.int/web/gaia/dpac/consortium}). Funding for the DPAC has been provided by national institutions, in particular, the institutions participating in the Gaia Multilateral Agreement. We thank the SIMBAD database and the online VizieR library service for helping us with the literature survey and obtaining relevant data.

\bibliography{AA_final}

\newpage
\appendix                  
\section{Additional data}

\begin{table}[!ht]%
    \centering
    \caption{The sample table of LAMOST LELS sources. }
    \begin{tabular}{|l|l|l|l|l|l|l|}
    \hline
        LAMOST desid & RA & DEC & LAMOST SpT & Classification & Group & Name \\ \hline
        LAMOST J080307.80+144949.5 & 120.782514 & 14.830419 & dM0 & dMe & B & LM-LEMC\_102 \\ \hline
        LAMOST J202431.01+421605.0 & 306.1292289 & 42.2680711 & G2 & CTTS & C & LM-LEMC\_10581 \\ \hline
        LAMOST J223304.19+234418.7 & 338.26746 & 23.738552 & K5 & Ke & C & LM-LEMC\_10960 \\ \hline
        LAMOST J013356.31+294201.1 & 23.484628 & 29.700313 & dM2 & dMe & B & LM-LEMC\_118 \\ \hline
        LAMOST J114044.45+310630.5 & 175.185241 & 31.108484 & K7 & Ke & B & LM-LEMC\_122 \\ \hline
        LAMOST J223227.51+303416.0 & 338.114653 & 30.571131 & dM5 & dMe & B & LM-LEMC\_13 \\ \hline
        LAMOST J181613.21+495205.2 & 274.0550637 & 49.8681296 & dM6 & dMe & A & LM-LEMC\_1464 \\ \hline
        LAMOST J110134.88+320928.8 & 165.395343 & 32.158016 & K7 & Ke & B & LM-LEMC\_149 \\ \hline
        LAMOST J111248.50+242320.4 & 168.202105 & 24.389021 & G7 & Ge & D & LM-LEMC\_171 \\ \hline
        LAMOST J043749.67+514255.8 & 69.456969 & 51.715519 & F0 & Fe & D & LM-LEMC\_201 \\ \hline
        LAMOST J121511.31+582749.5 & 183.797165 & 58.463768 & K7 & Ke & B & LM-LEMC\_205 \\ \hline
        LAMOST J035034.14+275836.5 & 57.642273 & 27.976813 & K7 & Ke & B & LM-LEMC\_21 \\ \hline
        LAMOST J052108.71-054229.0 & 80.286323 & -5.708056 & dM1 & CTTS & A & LM-LEMC\_2103 \\ \hline
        LAMOST J035603.23+351450.5 & 59.013466 & 35.247377 & dM1 & CTTS & A & LM-LEMC\_2211 \\ \hline
        LAMOST J101432.90+300616.4 & 153.637106 & 30.104568 & dM4 & dMe & B & LM-LEMC\_228 \\ \hline
        LAMOST J033233.00+310221.5 & 53.13753 & 31.039308 & gM6 & PtMS & A & LM-LEMC\_2516 \\ \hline
        LAMOST J063949.01+101602.8 & 99.954237 & 10.267467 & G5 & PtMS & D & LM-LEMC\_253 \\ \hline
        LAMOST J115740.34-011012.4 & 179.418112 & -1.170133 & dM5 & dMe & D & LM-LEMC\_267 \\ \hline
        LAMOST J063351.63+044942.8 & 98.465158 & 4.82856 & F5 & Fe & D & LM-LEMC\_316 \\ \hline
        LAMOST J061439.96-062917.4 & 93.666512 & -6.4881826 & K1 & Ke & A & LM-LEMC\_3273 \\ \hline
        LAMOST J062920.78+044006.9 & 97.336599 & 4.668605 & G3 & Ge & D & LM-LEMC\_375 \\ \hline
        LAMOST J221010.40+293625.8 & 332.543353 & 29.607178 & K7 & Ke & B & LM-LEMC\_4 \\ \hline
        LAMOST J063321.38+050018.8 & 98.339094 & 5.005244 & G5 & PtMS & D & LM-LEMC\_404 \\ \hline
        LAMOST J063416.41+044202.3 & 98.568391 & 4.700666 & F2 & Fe & D & LM-LEMC\_405 \\ \hline
        LAMOST J033344.06+073142.8 & 53.433604 & 7.528561 & dM0 & dMe & B & LM-LEMC\_41 \\ \hline
        LAMOST J063100.24+044738.2 & 97.751028 & 4.793951 & K5 & PtMS & D & LM-LEMC\_425 \\ \hline
        LAMOST J063003.35+044349.6 & 97.513998 & 4.7304482 & G6 & Ge & D & LM-LEMC\_439 \\ \hline
        LAMOST J004430.15+442703.6 & 11.125628 & 44.451011 & K7 & Ke & B & LM-LEMC\_526 \\ \hline
        LAMOST J005822.58+291420.3 & 14.5941057 & 29.238974 & K7 & Ke & B & LM-LEMC\_61 \\ \hline
        LAMOST J044049.50+255118.9 & 70.20628 & 25.855261 & dM2 & CTTS & B & LM-LEMC\_612 \\ \hline
        LAMOST J054112.83+252912.2 & 85.303487 & 25.486748 & dM4 & dMe & B & LM-LEMC\_647 \\ \hline
        LAMOST J025217.58+361648.1 & 43.073285 & 36.280052 & K4 & PtMS & A & LM-LEMC\_664 \\ \hline
        LAMOST J235556.66+324013.2 & 358.986106 & 32.670338 & F7 & Fe & D & LM-LEMC\_80 \\ \hline
        LAMOST J042951.55+260644.5 & 67.464831 & 26.112387 & dM0 & CTTS & A & LM-LEMC\_861 \\ \hline
        LAMOST J053232.08+104417.9 & 83.133682 & 10.738327 & dM4 & CTTS & A & LM-LEMC\_8805 \\ \hline
        LAMOST J034458.57+235541.0 & 56.244074 & 23.928062 & dM5 & dMe & A & LM-LEMC\_8860 \\ \hline
        LAMOST J053411.38-045122.9 & 83.547457 & -4.8563636 & dM4 & dMe & A & LM-LEMC\_9595 \\ \hline
    \end{tabular}
    \label{table:Catalog}
\end{table}



\label{lastpage}

\end{document}